%% file: paper.tex
\documentclass[sigconf,10pt]{acmart}

\usepackage{times}  
\usepackage{hyperref}
\usepackage{xcolor}
\usepackage{xspace}
\usepackage{subcaption}
\usepackage{enumitem}
\usepackage{caption}
\usepackage{amsmath}
\usepackage{etoolbox}
\usepackage{pgf}
\usepackage{adjustbox}
\usepackage{multirow}
\usepackage{tikz}
\usepackage{collcell}
\usepackage{hhline} 
\usepackage{tabularx}
\usepackage{float}
\usepackage{array}

\usepackage[bottom]{footmisc}   %
\usepackage[subtle]{savetrees}  %
\usepackage{multirow}
\newcommand{\myitem}[1]{\vspace*{0.02in}\noindent\textbf{#1}}
\usepackage[font=small,skip=8pt]{caption}
\def\showcomments{1}

\ifnum\showcomments=1
\newcommand{\aarti}[1]{{\footnotesize\color{magenta}[aarti: #1]}}
\newcommand{\fengchen}[1]{{\footnotesize\color{blue}[fengchen: #1]}}
\newcommand{\divya}[1]{{\footnotesize\color{green}[divya: #1]}}

\newcommand{\maria}[1]{{\footnotesize\color{red}[maria: #1]}}

\newcommand{\todo}[1]{{\footnotesize\color{green}[todo: #1]}}

\else
\newcommand{\fengchen}[1]{}
\newcommand{\divya}[1]{}
\newcommand{\maria}[1]{}
\newcommand{\aarti}[1]{}

\newcommand{\todo}[1]{}

\fi

\newcommand{\remove}[1]{}

\newcommand{\ie}{\emph{i.e.,}\xspace}
\newcommand{\eg}{\emph{e.g.,}\xspace}

\newcommand{\sys}{Zoom2Net\xspace}

\newcommand{\raw}{$T_{r}$\xspace}
\newcommand{\rawO}{$T_{r}^1$\xspace}
\newcommand{\rawT}{$T_{r}^2$\xspace}
\newcommand{\rawN}{$T_{r}^n$\xspace}
\newcommand{\rawI}{$T_{r}^i$\xspace}

\newcommand{\coarse}{$T_{s}$\xspace}
\newcommand{\coarseO}{$T_{s}^1$\xspace}
\newcommand{\coarseT}{$T_{s}^2$\xspace}
\newcommand{\coarseN}{$T_{s}^n$\xspace}
\newcommand{\coarseI}{$T_{s}^i$\xspace}

\newcommand{\collisions}{coarse-grained collisions\xspace}

\newcommand{\app}{$\hat{T}_{r}$\xspace}

\newcommand{\appI}{$\hat{T}_{r}^i$\xspace}

\newcommand{\de} {$S$\xspace}

\newcommand{\downstream} {{management}\xspace}

\renewcommand\footnotetextcopyrightpermission[1]{} %
\setcopyright{none}

\acmDOI{}

\acmISBN{}

\acmPrice{}

\begin{document}
\title{Super-resolution on network telemetry time series}

\settopmatter{authorsperrow=4}
\author{Fengchen Gong}
\affiliation{%
  \institution{Princeton University}
}

\author{Divya Raghunathan}
\affiliation{%
  \institution{Princeton University}
}

\author{Aarti Gupta}
\affiliation{%
  \institution{Princeton University}
}

\author{Maria Apostolaki}
\affiliation{%
  \institution{Princeton University}
}
\renewcommand{\shortauthors}{ }
\renewcommand{\shorttitle}{ }

\begin{abstract}
Fine-grained monitoring is crucial for multiple data-driven tasks such as debugging, provisioning, and securing networks. Yet, practical constraints in collecting, extracting, and storing data often force operators to use coarse-grained sampled monitoring, degrading the performance of the various tasks. 
In this work, we explore the feasibility of leveraging the correlations among coarse-grained time series to impute their fine-grained counterparts in software. 
We present \sys, a transformer-based model for network imputation that incorporates domain knowledge through operational and measurement constraints, ensuring that the imputed network telemetry time series are not only realistic but also align with existing measurements and are plausible. 
This approach enhances the capabilities of current monitoring infrastructures, allowing operators to gain more insights into system behaviors without the need for hardware upgrades.
We evaluate \sys on four diverse datasets (\eg cloud telemetry and Internet data transfer) and use cases (such as bursts analysis and traffic classification).
We demonstrate that \sys consistently achieves high imputation accuracy with a zoom-in factor of up to 100 and performs better on downstream tasks compared to baselines by an average of 38\%.

\end{abstract}

\maketitle
\input{1_intro}
\input{3_motivation}
\input{4_overview}
\input{design}
\input{eval}
\input{conclusion}

\bibliographystyle{ACM-Reference-Format}
\bibliography{reference}

\end{document}

%% file: 1_intro.tex
\section{Introduction}

Imagine a large datacenter operator tasked with pinpointing the root cause of an instance of packet drops occurring on a network switch. They speculate on several potential issues, such as buffer shortage, bursty traffic, misconfiguration, or potential hardware problems. To identify the actual cause, the operator examines different monitored signals, including active end-to-end latency, packet counts, and queue lengths. However, they quickly realize that the collection interval of these measurements is not fine-grained enough to provide the necessary insights. This is a typical problem as very fine-grained monitoring is hindered by hardware limitations or cost factors.
In theory, the operator could request the collection of more fine-grained queue-length data from all devices and wait for the next instances of packet drops to identify the root cause. 

As collecting more fine-grained telemetry will be time-consuming and even infeasible for some devices, the operator would, most likely, end up looking back at the existing coarse-grained measurements, trying to piece together likely scenarios in their mind. This involves speculating about concrete scenarios, mentally constructing a finer-grained version of the measured signals, and aligning these speculations with the actual measurements they have. 
For instance, if there were spikes in end-to-end latency measurements of distinct queues before the spurious drops, the operator might speculate that multiple queues' lengths were consistently high, which allowed the random sampling to `catch' multiple high values, then those consistently high queues could have filled the switch buffer, leading to drops. 
This reasoning is feasible due to the underlying correlations between various network signals (\eg end-to-end delay, queue lengths, drops), which the operator understands and could use to make informed inferences (guesses) about the events that occurred.

While intuitive in this example, conceptualizing various signals and filling the gaps in monitoring generally is extremely challenging even for seasoned operators.
The challenge lies in accurately identifying, utilizing, and making sense of the correlations among various signals.  Additionally, the sheer breadth of the potential search space adds to the complexity, making both manual and automated reasoning methods ineffective due to scalability concerns. %
Still, the scenario raises a question: can we design a system that will automatically analyze multiple correlated, coarse-grained network signals to recover a more detailed picture of the network?  
Doing so would allow operators to maximize the value of their existing monitoring infrastructure, indirectly improving multiple management tasks with no hardware investment.  

We introduce \sys, a system that imputes fine-grained network monitoring data from multiple coarse-grained ones. 
This is possible because the various sampled (coarse-grained) time series not only constrain their own imputed versions
(imputed signals need to be consistent with the measurements)
but also 
impose constraints on
the imputed versions of each other. 
For instance, coarse-grained queue-length samples constrain the imputed, more fine-grained queue lengths, and these, in turn, are also constrained by the packet counts
(as a queue requires a sufficient number of packets to be received to form).
While \sys addresses an inherently under-constrained problem, we observe that there are additional correlations among measured time series that further narrow down potential outputs, often associated with traffic patterns or challenging-to-formulate correlations. 
Under these correlations, certain scenarios are more likely to occur repeatedly, contributing to a more predictable aspect of the problem.
\sys can explore the correlations within a small set of fine-grained data generated from an expensive on-demand monitoring tool (or small packet trace) and then be used to improve the granularity of always-on monitoring infrastructure.

At first glance, \sys resembles super-resolution, which recovers high-resolution images from their low-resolution counterparts using deep learning: a concept extensively explored in the literature~\cite{ledig2017photo,wang2018esrgan,wang2021real,zhang2020deep, zhang2018residual,dong2015image}. 
Similarly to images where there are correlations within the RGB values and across adjacent pixels
allowing the imputation tasks, network signals are often correlated. Sometimes, those correlations can be expressed as mathematical equations or limits,  while other times, they manifest as hidden patterns.

Despite the rich literature, off-the-shelf ML models~\cite{cao2018brits} prove inadequate for \sys. 
Network telemetry imputation demands not only realism but also consistency with existing measurements and adherence to known principles, a requirement often 
neglected by general ML models.
Unlike typical super-resolution tasks, which prioritize visually appealing outputs, network imputation necessitates
recovered fine-grained time series to \emph{(i)} closely resemble ground truth; and \emph{(ii)} produce the original coarse-grained time series if sampled with the same sampling process.
Integrating this kind of domain knowledge into ML models is complex: designed
to learn from data, ML models cannot easily ingest traditional
knowledge, such as rules and relationships. As a result, there is no standard way of doing so.
Moreover, defining the success criteria for network imputation proves challenging. Commonly used metrics like mean-square error (MSE) can penalize outputs that are practically equivalent, such as temporally shifted bursts. Even more complex is the scenario where the same coarse-grained data corresponds to multiple fine-grained possibilities in the training set (ambiguity), potentially hindering the ML model's effective training.

Instead of 
solely relying on data for training, \sys incorporates different sources of knowledge to generate not only realistic but accurate fine-grained time series using two specific methods.
First, utilizing a transformer model~\cite{vaswani2017attention} at its core, \sys employs a knowledge-augmented loss function that embeds both \emph{operational} and \emph{measurement constraints}, guiding the transformer 
to learn both correlations and properties. 
Second, \sys enforces the constraints that the transformer failed to satisfy by using a \emph{constraint enforcement module} to correct %
its output post-imputation.

To manage the inherent ambiguities in data, which tend to increase as the zoom-in factor (the ratio between coarse and fine granularity) becomes more pronounced, we have modified our strategy. Instead of aiming for precise, point-to-point accuracy regardless of the zoom-in factor, we train \sys to generate outputs that are plausible, consistent with measurements, and functionally equivalent to the ground truth. This change not only improves the efficiency of the training process but also leads to the production of outputs that are more practical and useful in real-world applications.

We evaluate \sys across synthetic and real-world datasets and compare its performance to both statistics and learning-based methods. We find \sys:
\emph{(i)} consistently achieves high imputation accuracy with a zoom-in factor of up to 100;
\emph{(ii)}  achieves 38\% better performance in downstream tasks compared to baselines; and
\emph{(iii)}  demonstrates its ability to apply learned correlations to scenarios unseen during training.

%% file: 3_motivation.tex
\section{Motivation and limitations of existing work}
In this section, we discuss motivating scenarios where fine-grained telemetry is required after the fact. Next, we explain why existing solutions are unlikely to help.

\subsection{Use cases and requirements}
\label{sec:usecase}

\myitem{Post-Mortem Analysis.} %
In the aftermath of a volumetric attack, network operators leverage monitoring to analyze the event. This data helps in revealing the attack's impact across the network, pinpointing vulnerable zones, and assessing the defense mechanisms' performance under real stress. 
Insights gained from this analysis are crucial for strengthening the network's resilience against future attacks. They can guide adjustments in security policies, firewall rules, and even the deployment of advanced intrusion detection systems that can better handle sudden surges in traffic. 
While vital, the collection and storage of every potential signal at the finest level of granularity continuously is impractical.
Thus, having a way to zoom into a particular signal that is being monitored after the fact is extremely useful. 

Intuitively, a learning approach is unlikely to help in identifying a novel attack. Yet, in the context of network behavior, individual monitoring signals may not deviate significantly from normal patterns.
In other words, a novel or rare attack could manifest as an escalation in the frequency of regular incidents, or it may be the result of a specific sequence and combination of routine occurrences that serve the attacker's objectives rather than a set of anomalous or unforeseen behaviors on each signal. Hence, a learning approach can, in fact, zoom into the incident. 
For example, consider a shrew attack~\cite{shrew} characterized by periodic bursts that disrupt TCP. 
Such malicious behavior can only be revealed by \emph{fine-grained} queue lengths; each 
burst is a regular incident commonly seen.

\myitem{Diagnostic Troubleshooting.} %
To identify the root cause of an alert or client's complaint, network operators need to check various monitored signals potentially on multiple devices and try to identify anomalies or deviations from normal operations.
 Note that these anomalies could only be visible when the monitoring interval is adequately small, while the exact signals of interest are not necessarily known. For instance, dropped packets in a flow caused by bursts can only be observed with millisecond granularity measurements on all devices on each of the paths through which a flow could have been forwarded. 
Thus, fine-grained and general monitoring is critical for prompt and precise troubleshooting and, thus, faster resolution of network issues. We evaluate this use case in \S\ref{sec:case1}.

\myitem{Network resource planning.}
To plan for capacity increases, on-device memory provisioning, topology design, peering agreements, and many %
other resources, an operator needs detailed insights into the network's usage patterns and demands over long periods of time. 
Fine-grained monitoring of various signals would help to avoid both under-provisioning, which can lead to congestion and performance issues, and over-provisioning, which can be costly and inefficient. On the one hand, collecting and storing all possible monitoring data for long periods to satisfy future operators' demands is wasteful, given the data volume. On the other hand, predicting signals of interest and their required granularity is impractical given the technological advancements and evolving application demands.
For example, an operator might need to decide on the buffer size to provision on network devices, for which they would need to see historical data on the aggregate occupancy of the buffer in quantity (percentage of buffer occupied) and in quality (stably long queues or bursty traffic). Yet, the data might have been collected at a time when buffer size was not a limiting resource, hence not a signal of interest, or at a time when the latency of milliseconds was acceptable; hence, higher granularity seemed unnecessary. In such cases having a way to zoom into signals would be a great software alternative. We evaluate this use case in \S\ref{sec:case1}, and \S\ref{sec:case3}.

\myitem{Dataset imputation.}
To protect their networks from attacks, load balance traffic, or debug issues, operators often use machine learning. To train their models, the operators will turn the initial packet traces into a series of features. Oftentimes, the raw traces are deleted after the features have been calculated to reduce the risks of a data breach or to save space. The features are stored and/or shared. While convenient, this solution prevents the operators from going back to the raw trace and calculating new features that they conjecture might be useful for their operation. Of course, they might be able to collect new traces if they own the infrastructure, but that would be time-consuming and might hurt performance (e.g., if the previous dataset happened to contain attacks or infrequent scenarios). In other words, many datasets today can be seen as coarse-grained representations of raw data, retrieving which would allow for easier experimentation. For example, one might have collected a trace of website accesses from a developing country. They calculated the useful features before sending the dataset to reduce the resources needed for transfer and storage. Yet they soon realized that there was one feature that they had missed. In this case having a way to `zoom into' features to find the raw dataset would be very useful. While this use case might deviate from our core goals, we see encouraging results in \S\ref{sec:case4}; after all, feature engineering on network traces is a form of sampling.

\subsection{Limitations of existing works}
\label{sec:requirement}
In various use cases, a recurring theme is the need for techniques that gather telemetry data to meet three specific requirements:
\emph{(i)} \textbf{generality}: the technique should be applicable to many different types of telemetry. This is critical as
operators are often uncertain about signals of interest \emph{a priori}; 
\emph{(ii)}  \textbf{fine granularity}:  telemetry should be collected frequently enough to facilitate the detection of short-timescale changes (\eg bursts); %
 and 
\emph{(iii)} \textbf{ cost efficiency}: the processing, memory and storage needed for telemetry should be low and amenable to be deployed on commodity hardware. 
Next, we summarize why existing monitoring solutions failed fulfill these requirements.

\myitem{Traffic mirroring: } Mirroring traffic from routers to collectors ~\cite{tilmans2018stroboscope,rasley2014planck}  using 
specialized hardware can satisfy granularity and generality as they may allow an operator to run custom queries at any granularity. 
However, this does not meet cost efficiency as traffic volumes increase; \eg mirroring requires significant bandwidth and processing, and custom hardware may become too expensive. 

\myitem{Sampling \& Network Tomography: } Packet sampling approaches (\eg SNMP, sFlow) can reduce costs. Unfortunately, polling counters is resource intensive and hence typically done at a coarse time-scale, \eg minutes. Thus, it is not fine-grained and cannot be used to detect microbursts~\cite{teixeira2008impact}. Network tomography
can make more out of the collected data, not by increasing the granularity but by inferring unmeasured metrics (\eg latencies of sub-paths). Both traditional and more advanced network tomography~\cite{geng2019simon} only work on linear relations between signals and known network topology, lacking generalization.

\myitem{Programmable switch \& eBPF: }
Recent developments in programmable switches have enabled different methods of collecting network telemetry. 
In-band network telemetry (INT) is a technique that embeds fine-grained and accurate telemetry in each packet. While effective, INT generates a substantial amount of data, leading to high memory and bandwidth usage, thus, is not cost-efficient. Lighter versions of INT ~\cite{ben2020pint,qian2023offset,langlet2023dta} still require homogeneous and programmable hardware and generate a large amount of data.
Customized algorithms such as sketches offer a cost-efficient approach to generating fine-grained telemetry.
However, these algorithms are typically designed for specific tasks at design time (\eg heavy-hitter detection~\cite{li2016flowradar}, RTT monitoring \cite{RouteScout, sengupta2022continuous}) or identifying culprits ~\cite{lei2022printq,chen2019conquest}, lack generalization across tasks, and often generate huge amounts of data that are hard to extract from the device and store. 
Even highly optimized solutions using eBPF often fall into scalability walls. For instance, to reduce its memory and CPU usage, Millisampler~\cite{ghab2022millisamp} can only monitor and collect fine-grained telemetry for a short period (\eg 20 seconds), potentially missing the most critical events. 

%% file: 4_overview.tex
\section{Overview}
Traditional monitoring techniques fall short in one or more of the requirements (cost-efficiency, fine granularity, and generality) in \S\ref{sec:requirement}. 
Our overarching goal is to develop a framework that can fulfill these requirements in software through network telemetry imputation. We formulate the problem and then explain the challenges and insights that drove our design.

\begin{figure}[t!]
        \centering
\includegraphics[width=0.7\linewidth]{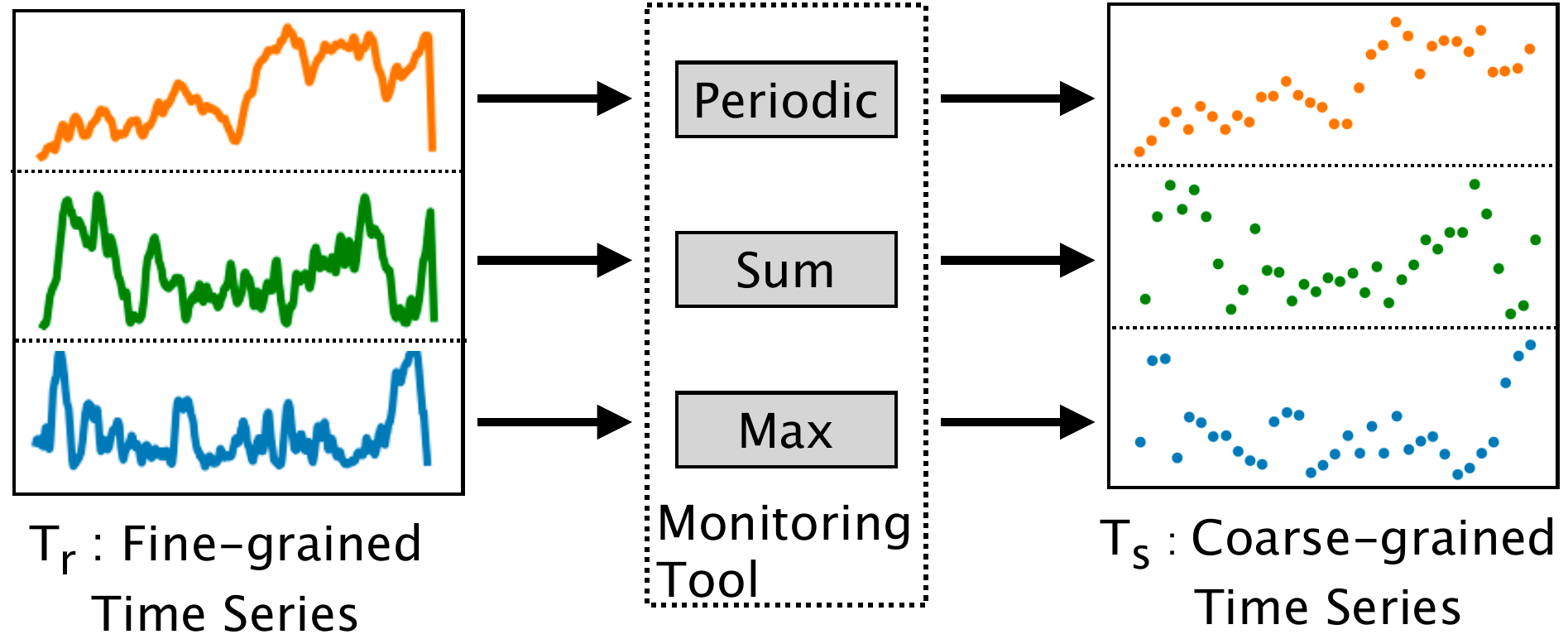}
\captionsetup{skip=3pt}
    \caption{Fine-grained signals \raw of a networked system are sampled by monitoring tools, resulting in coarse-grained time series \coarse available to operators. In practice, \raw is only available for training and is collected over a short period using specialized hardware or traffic mirroring.\vspace{-1em}}
    \label{fig:sampling}
\end{figure}

\subsection{Network imputation}
\begin{figure}[t!]
        \centering
\includegraphics[width=\linewidth]{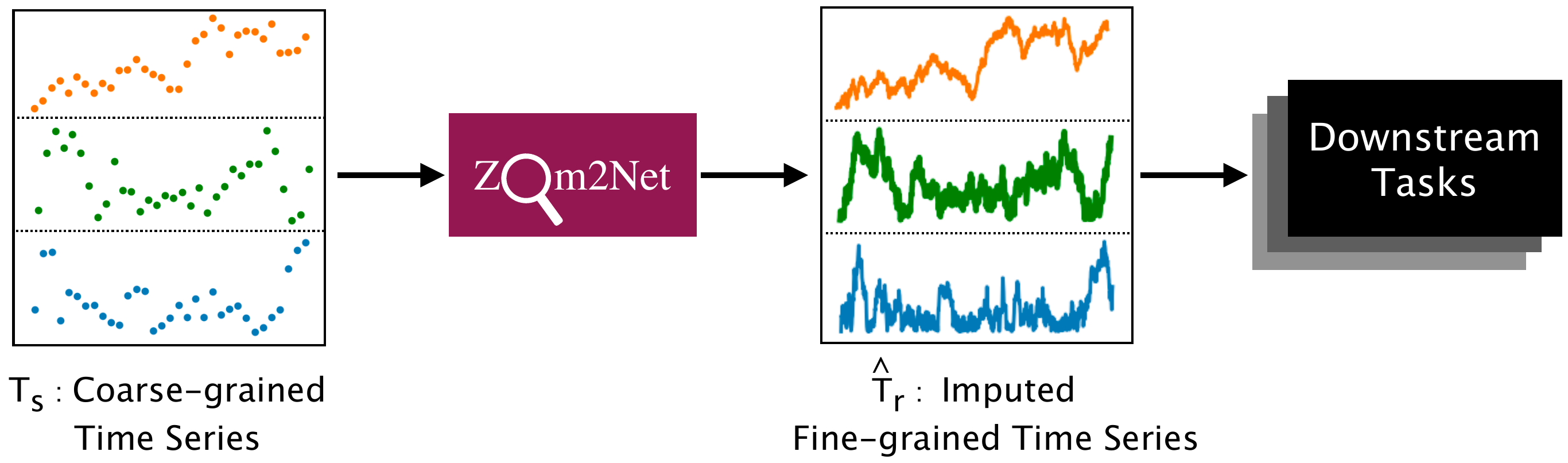}
\captionsetup{skip=3pt}
    \caption{\sys takes a set of coarse-grained time series \coarse as input and outputs imputed fine-grained time series \app which is fed to multiple downstream tasks. \vspace{-1em}}
    \label{fig:impute}
\end{figure}

Consider a physical networked system that is 
described by a set of fine-grained \emph{raw} time series \raw = \{\rawO, \rawT,...\rawN\}. 
A monitoring infrastructure senses the physical system using a known sampling/coarsening function \de, and outputs \coarseI= \de (\rawI). 
For instance, \de could sample a single value out of every $N$ values in \rawI, or it could output the average/max/mean for every window. 
In total, the output of the sampling function is a set of coarse-grained time series \coarse = \{\coarseO, \coarseT,...\coarseN\}. 
We illustrate this in Fig.~\ref{fig:sampling}.
We aim to recover (impute) an approximate \rawI: \appI,  
which, if given as input to the various \downstream tasks, can improve their performance compared to if their input was from \coarseI. This process is illustrated in Fig.~\ref{fig:impute}.
\raw is sampled above the Nyquist frequency; thus, there is information loss that prevents its trivial recovery. 
We assume there is available a dataset of fine-grained time series from the network of interest. This can be practically generated by running for a short period: \emph{(i)} an unscalable monitoring tool that cannot be always-on because of CPU/storage usage, or \emph{(ii)} an advanced hardware device only temporarily plugged into the network or \emph{(iii)} a simple tab/mirror of traffic.
We do not assume the \coarse to be perfectly aligned in time nor monitored on the same granularity.

\myitem{Non goals.}
Our system works offline for model inference and data analysis.
While this would prevent real-time tasks, our system could already be used to improve debugging, provisioning, and attack analysis (including all the use cases in \S\ref{sec:usecase}).

\subsection{Challenges and Insights}
\label{sec:challenge}
This section describes the challenges and insights.

\myitem{Challenge 1: A single coarse-grained time series is often ambiguous, making imputation impossible.}
The network imputation problem is, by nature, under-constrained: 
multiple versions of fine-grained time series, when sampled \eg below the Nyquist rate, can produce identical coarse-grained time series. As a result, it is fundamentally difficult (if not impossible) to reconstruct the correct fine-grained version.

\begin{figure}[t!]
    \centering
    \begin{subfigure}[b]{0.9\linewidth}
        \centering
        \includegraphics[width=0.7\linewidth]{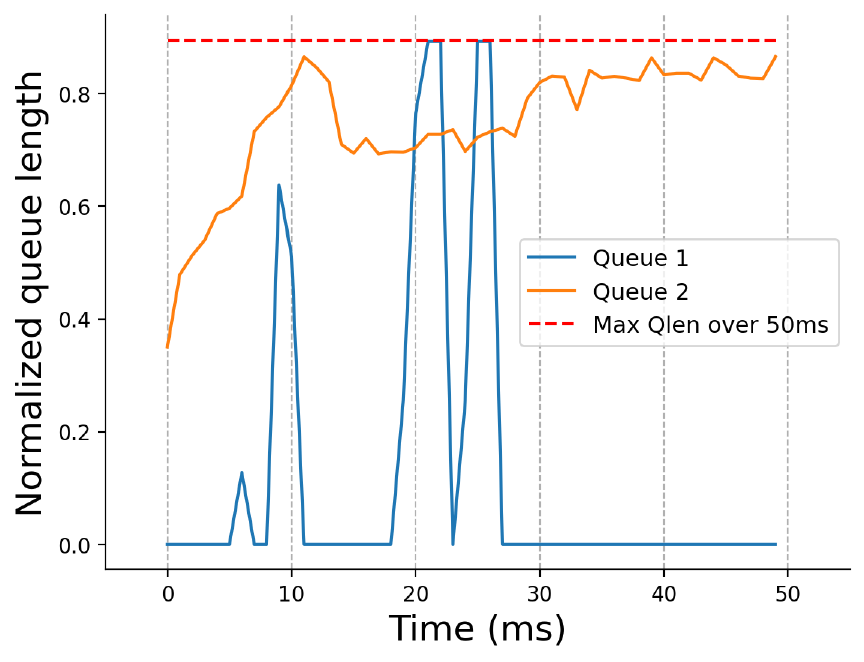}
        \caption{Two fine-grained queue length time series with distinct behaviors at 1ms.}
        \label{fig:same-max}
    \end{subfigure}
    \begin{subfigure}[b]{\linewidth}
        \centering
        \begin{tabular}{ |c|c|c|c| }
    \hline
    & Max Qlen &  Packet Drop & Packet Sent \\ \hline
    Queue 1 & 0.895 & 0.05 & 0.74 \\ \hline
    Queue 2 & 0.87  & 0.69 & 0.98 \\ \hline
    \end{tabular}
        \caption{Signals for two queues sampled at 50ms.}
        \label{tab:queue_result1}
    \end{subfigure}
    \captionsetup{skip=3pt}
    \caption{Two distinct fine-grained queue length behaviors result in the same sampled maximum queue length. But they can be distinguished from different sampled packet drops and sent counts.\vspace{-1em}} 
    \vspace{-0.1in}
\end{figure}

We show an example in Fig.~\ref{fig:same-max}, where we plot the queue length of two queues at fine-grained intervals (at 1ms).
Queue 1 has multiple short bursts while queue 2 has a constantly high queue length, despite them showing close maximums when sampled every 50ms. 
While the queues display distinct patterns, the result of reasonable monitoring tools such as LANZ~\cite{lanz}, which outputs the maximum queue length at every interval (50ms in this case), results in coarse-grained time series that are identical. 
This is problematic for an operator observing only the coarse-grained version as these different queue patterns necessitate varied approaches for resolution. For instance, a consistently high queue, such as queue 2 (orange), may require adjustments in congestion control or the adoption of an Active Queue Management (AQM) system. Conversely, a bursty queue pattern, such as queue 1 (blue), suggests the presence of a bursty application, calling for a strategy like pacing for effective management.

\myitem{Insight: Leveraging multiple time series can often resolve the ambiguity.}
Relying solely on a single coarse-grained queue monitoring time series may not suffice to impute fine-grained queue length. Still, integrating additional time series can provide the necessary clarity.
Specifically, the two queues in Fig.~\ref{fig:same-max}  naturally exhibit significantly different patterns in terms of packet counts and drop counts which are also typically monitored \eg by SNMP~\cite{snmp1}. We can observe this in the normalized Packet drop and Packet sent columns of Table~\ref{tab:queue_result1}. Thus, by examining SNMP-like data covering the same time interval, we can discern distinct traffic profiles for these two queues.  In other words, because queue lengths, packet counts and drop counts time series are correlated, we can achieve higher accuracy in imputing fine-grained queue length if we use all three coarse-grained signals compared to using only maximum queue length.

This situation is typical in networking, where various monitored time series often display correlations.
For example, in data centers, traffic rates for servers within the same Top-of-Rack (ToR) are correlated because they share an uplink (thus, these servers cannot be concurrently sending data at full capacity in oversubscribed networks).
Similarly, end-host traffic volume correlates with congestion window size and Round Trip Time (RTT), which denote current network conditions.

Leveraging these correlations is a great opportunity for solving the imputation problem, but they are not trivial to capture. 
Recent advancements in generative models~\cite{lin2020dp,yin2022netshare,buehler2020data,brown2020gpt,jiang2023diffusion}, offer a promising solution. Transformers, as sequence-to-sequence models, excel at learning correlations across lengthy sequences through parallel processing via the attention mechanism~\cite{vaswani2017attention}. Their capacity to generalize effectively and capture diverse contexts allows them to grasp shared dynamics without overfitting. This flexibility and efficiency have already established their popularity in the field of networking~\cite{dietmuller2022transformer, le2022foundation, houidi2022represent}.

\begin{figure}[t!]
    \centering
    \begin{subfigure}[b]{0.9\linewidth}
        \centering
        \includegraphics[width=0.7\linewidth]{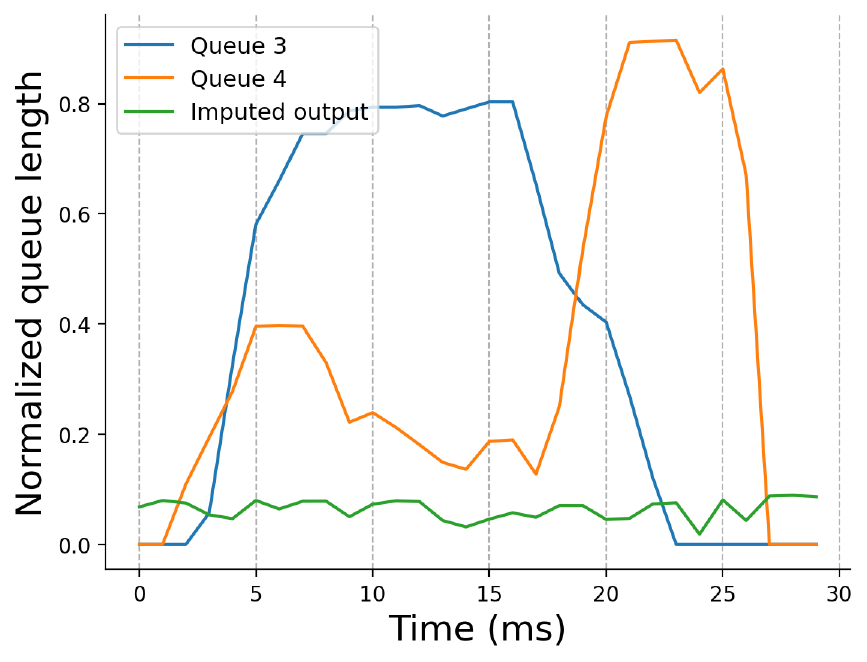}
        \caption{Two fine-grained queue lengths at 1ms with an imputed output from a transformer model.}
        \label{fig:same-all}
    \end{subfigure}
    \begin{subfigure}[b]{\linewidth}
        \centering
        \begin{tabular}{ |c|c|c|c| }
    \hline
    & Max Qlen &  Packet Drop & Packet Sent \\ \hline
    Queue 3 & 0.80 & 0.06 & 0.60 \\ \hline
    Queue 4 & 0.92  & 0.086 & 0.64 \\ \hline
    \end{tabular}
        \caption{Signals for the two queues sampled at 50ms.}
        \label{tab:queue_result2}
    \end{subfigure}
    \captionsetup{skip=3pt}
    \caption{Two distinct fine-grained queue length encompass almost identical coarse-grained signals. A transformer model trained with MSE, having seen both (and more), would generate an average (green line), obfuscating the burst.\vspace{-1em}}
    \vspace{-0.15in}
\end{figure}

\myitem{Challenge 2: Ambiguity might remain even when combining multiple coarse-grained time series.}
While some scenarios can be recovered correctly by leveraging multiple correlated coarse-grained time series, that is not always the case.
Indeed, as we decrease the granularity of the coarse data (increase the zoom-in factor), there will be more clusters of pairs (\raw, \coarse) with identical coarse-grained series (\coarse) but distinct fine-grained counterparts (\raw). 
Fig.~\ref{fig:same-all} illustrates such a case with two queues experiencing a burst at different times, hence having very different fine-grained queue length time series.
Unlike the previous case (Fig.~\ref{fig:same-max}), though, all coarse-grained time series \coarse, namely the maximum queue length, the packet drop counts, and the sent packet count, are almost identical, as we observe in Table~\ref{tab:queue_result2}. 
Such cases, which we call \collisions, are detrimental to both training and inference.
First, a model trained on a dataset with multiple \collisions would take longer to converge and might be unstable. At a high level, this is the same problem as having a sample with multiple labels~\cite{wu2023learning, wu2022label}.  
Second, the output of an ML model in  \collisions might be useless. To illustrate this problem, we train a transformer with simple MSE loss and observe its behavior in imputing queue 3.
The transformer ends up producing the average of all scenarios with the same coarse-grained input, as we illustrate in Fig.~\ref{fig:same-all} in green, which would be completely useless (as it hides the burst itself). 

\myitem{Insight: A functionally equivalent imputation output is often a more attainable and useful goal than a perfect one.} 
We observe that oftentimes in \collisions \ie when all correlated coarse-grained time series are similar, the corresponding ground-truth fine-grained time series (albeit seemingly different) correspond to a functionally \emph{equivalent} scenario. In our example in Fig.~\ref{fig:same-all}, for instance, there is a burst of similar duration and rate that is shifted in time.  In such an instance, a system generating any of these equivalent versions would generally meet the expectations of the network operator. On the contrary, 
an average of multiple instances(as we illustrate in Fig.~\ref{fig:same-all}) would not be acceptable.
Ultimately, our primary goal was to automate the thought process of a highly skilled operator, who would typically attempt to align the coarse-grained data with anything they have seen before and matches.
Nonetheless, determining the equivalence of different fine-grained time series can be complex and might even depend on the specific downstream task for which the data is used. 

\begin{figure}[t!]
        \centering
\includegraphics[width=0.7\linewidth]{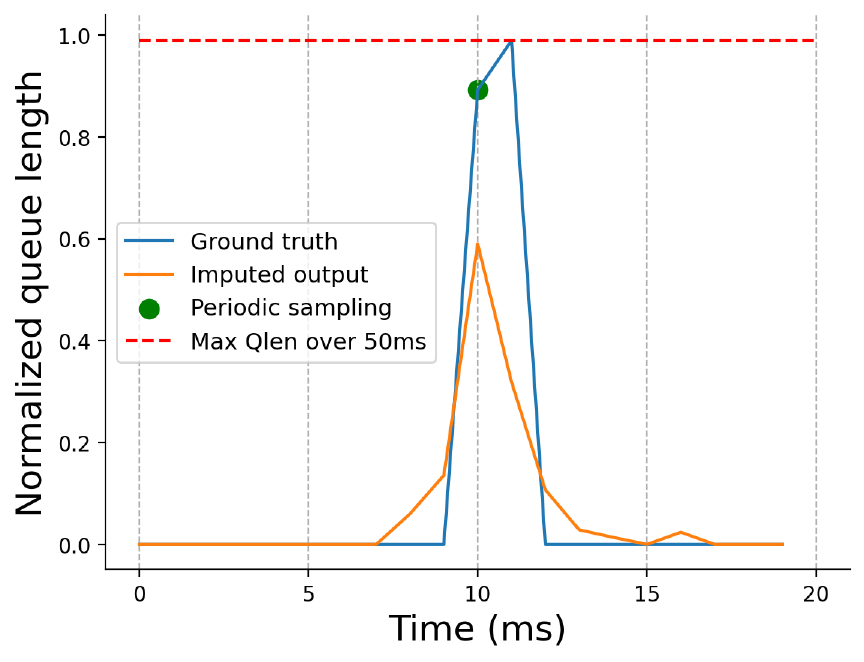}
\captionsetup{skip=3pt}
    \caption{Ground truth fine-grained queue length at 1ms (blue) and imputed fine-grained queue length at 1ms (orange).
    A plain transformer catches trends but outputs results that are inconsistent with maximum and periodic samples.\vspace{-1em}}
    \label{fig:violate}
\end{figure}

\myitem{Challenge 3: ML does not provide guarantees.}
The most significant downside of any ML in the context of telemetry imputation is that the output lacks correctness guarantees.
For example, in Fig.~\ref{fig:violate} we observe that the imputed time series (orange line) of a queue's length (blue line) generated by a transformer with MSE loss (plain transformer)
is not consistent with the measurements: the transformer did not impute a queue length that is as high as the (known) max queue length (red dashed line) of the interval, and the output at 10ms is not consistent with the corresponding periodic sample, although they are part of the transformer's input. 
It may seem surprising that the model does not "realize" the connection between the provided max/periodic queue lengths and the ground-truth (fine-grained) queue lengths. However, this issue arises due to the inherent challenge of predicting large values when the input data is predominately skewed towards smaller values.
To make matters worse, the output of our model also violated switch-specific constraints. For example, the total number of packets that would need to have been dequeued for the imputed queue to be formed exceeded the SNMP count.

\myitem{Insight: Incorporate knowledge and enforce consistency with measurements.}
Instead of solely relying on data to train our model, we leverage knowledge. Concretely, we observe that there are often correlations across monitored time series that can be formalized. For instance, given a queue length time series, we can calculate the number of packets dequeued, which should not exceed the number of packets sent out. Further, we know that the system's output should be consistent \ie produce the coarse-grained time series if sampled with the corresponding operator. We can leverage these connections to guide the output towards more accurate results. 
Inspired by Physics and the extensive work in Physics-Informed-Neural-Networks~\cite{pinn,djeumou2022neural} and the imaging literature~\cite{bahat2020explorable}, we incorporate this knowledge during training on the loss function and during inference through a consistency enforcement module.
By embedding domain knowledge,
we increase not only the model's accuracy but also its reliability \ie the operator can have more confidence that the result produced is plausible.

%% file: design.tex
\section{\sys Design}
\begin{figure}[t!]
        \centering
\includegraphics[width=\linewidth]{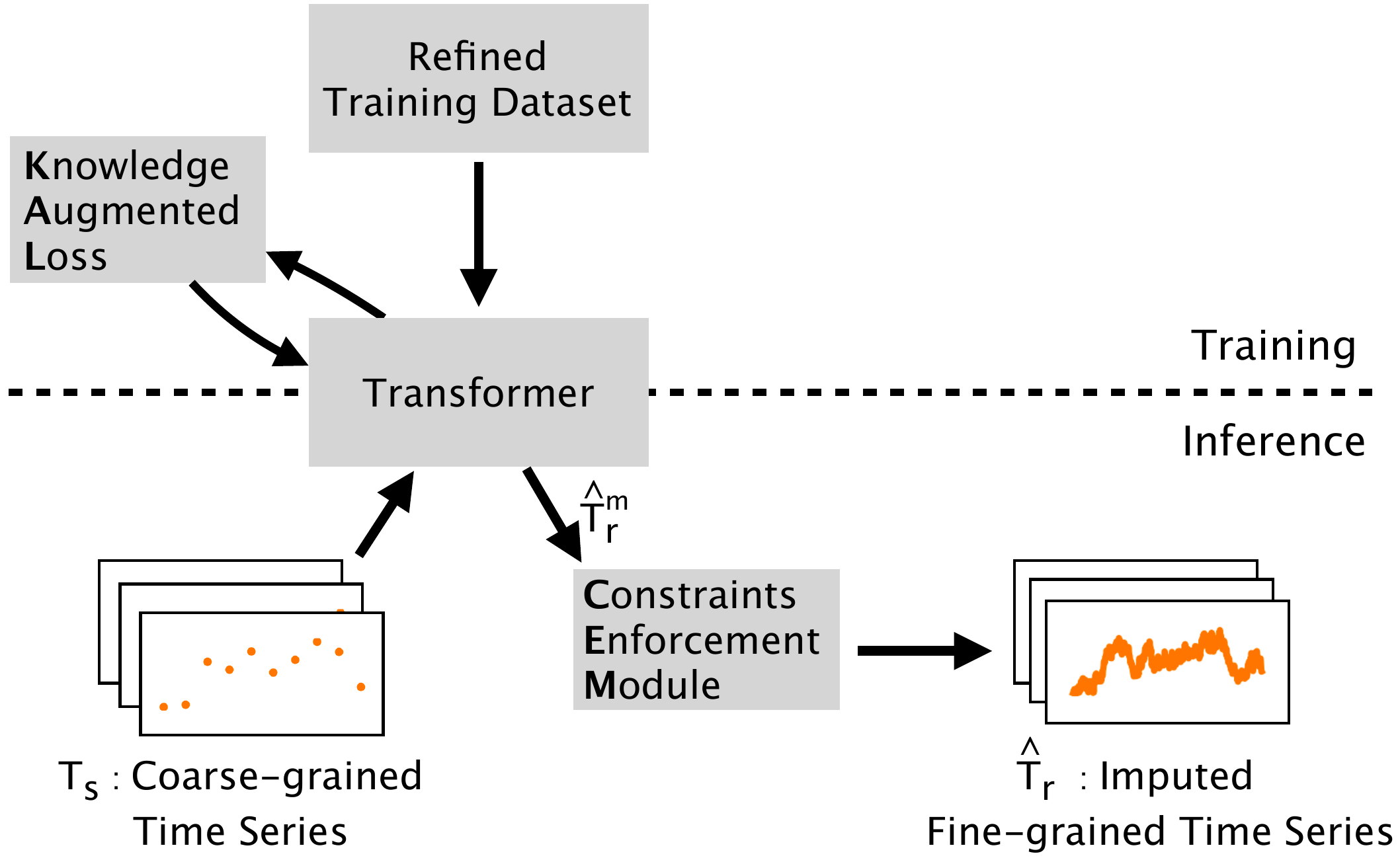}
\captionsetup{skip=3pt}
    \caption{During training, a transformer model takes refined training dataset and trains with Knowledge Augmented Loss function. During inference, the model takes in coarse-grained time series \coarse and outputs $\hat{T_r^m}$ which is then corrected by Constraints Enforcement Module. The result \app is used for downstream tasks. \vspace{-1em} }
    \label{fig:diagram}
\vspace{-0.04in}
\end{figure}

Driven by these insights, we design \sys, which we illustrate in Fig~\ref{fig:diagram}.
We start with discussing the formulation of knowledge (\S\ref{sec:formulation}), which we incorporate into the loss function, forming the Knowledge Augmented Loss (KAL) function (\S\ref{sec:kal}), and into a post-imputation Constraints Enforcement Module (CEM) for output correction (\S\ref{sec:cem}). Finally, we discuss ways of refining our training dataset to reduce the effect of \collisions (\S\ref{sec:refine}). 

\subsection{Knowledge formulation}
\label{sec:formulation}
In the context of \sys, we categorize knowledge into two types: measurement knowledge and operational knowledge. 

Measurement knowledge demands that 
applying monitoring tools (\eg max queue length, packet counts) on the imputed fine-grained time series output should result in coarse-grained measurements.
Inspired by formal methods, we articulate such knowledge as equality constraints $\Phi(\hat{T_{r}},T_{s})$.
\begin{align*}
\tag{$C_{equal}$}
    \Phi(\hat{T_{r}},T_{s}) & = 0  \label{eq:constr_eqal}\\
 where \quad
 \Phi(\hat{T_{r}},T_{s}) & = T_{s} - S(\hat{T_{r}})
\end{align*}
Here, $S$ represents the sampling/coarsening function.
Importantly, such constraints are extremely easy to identify because they are the result of monitoring.

The operational knowledge captures the correlations between different signals. For instance, the count of enqueued packets should not exceed the count of sent packets.
These relationships are expressed as inequality constraints $\Psi(\hat{T_{r}},T_{s})$.
\begin{equation*}
\tag{$C_{inequal}$}
    \Psi(\hat{T_{r}},T_{s}) \le 0 
\label{eq:constr_ineqal}
\end{equation*}
There are cases where correlations include comparison and logical operators (\eg $\leq$, $\lor$).
To integrate such constraints into our system, we transform the operators into expressions that output $True$ or $False$. 
For example, to test if the constraint $a < b$ is satisfied, we formulate it as $step(a-b)$ with a step function.

 Identifying and formulating measurement constraints is straightforward, as operators are typically familiar with their monitoring functions/tools. Formulating operational constraints might require more domain knowledge, but identifying them can be automated by running pure correlation tests \eg Pearson's Correlation Coefficient among the available signals.

\sys does not necessitate time series to be perfectly aligned. Indeed, as the transformer learns directly from data it can catch correlations from time series that are not perfectly synchronized (\eg due to unsynchronized clocks or because of different granularity intervals ). If there is an unknown time shift, it is (in theory) possible that the operational constraints are violated in ground truth (thus better to not be used). Yet, in practice, we don't see this because operational constraints are typically loose. Measurement constraints are not affected.

\subsection{Knowledge Augmented Loss (KAL)}
\label{sec:kal}
We designed our loss function to enable the model to learn data correlations and distributions, while also guiding it to satisfy the constraints derived from our knowledge.

\subsubsection{Loss metrics}
Loss functions that minimize point-wise distance, such as MSE, are designed to make the output closely resemble the ground truth, allowing the model to learn data correlations and patterns.
However, we noted that MSE encourages the model to find averages among plausible solutions. 
This becomes particularly problematic when the data is predominately skewed towards smaller values,
resulting in overly smooth outputs that struggle to capture bursts~\cite{ledig2017photo}.
To address the challenge of preserving correlations while overcoming the tendency towards averaged behaviors,
we introduce Earth Mover's Distance (EMD) as another term in the loss function.
EMD measures the minimum cost required to transport the mass of one distribution to match another. 
By incorporating EMD, we aim to
encourage the model output not only to closely match target values but also to mirror the structure and uncertainty inherent in the target distribution. 
Furthermore, EMD can be applied without considering the distance between masses. 
This addition alleviates the issue of \collisions because it naturally identifies the closeness of functionally equivalent outputs.
For instance, in terms of EMD, the two bursts in Fig.~\ref{fig:same-all} have similar fine-grained queue lengths and thus are not even \collisions.   
By minimizing both MSE and EMD, our loss function guides the model to adapt to the data's characteristics.
This combined loss function is expressed as:
\begin{equation*}
    L_{combine} = MSE(\hat{T}_{r}^m, T_{r}) + \lambda \ EMD(\hat{T}_{r}^m, T_{r})
\end{equation*}
where $\hat{T}_{r}^m$ is model output and $T_{r}$ is target value, $\lambda$ is a hyperparameter to balance the two loss terms.

\subsubsection{Knowledge constraints satisfaction}
Given the equality and inequality constraints defined in \S\ref{sec:formulation}, the challenge is how to inform the transformer of the knowledge we have. 
To address this limitation, our goal is to solve for a set of transformer parameters minimizing the loss $L_{combine}$ over the training dataset while also satisfying all the knowledge-based constraints. 
That is, we aim to solve the optimization problem:
\begin{align*}
\tag{1}
    \min\ L_{combine} \; \textrm{s.t.}\; & \Phi_k(\hat{T_{r}^m},T_{s}) = 0, \quad k \in \{1,...,K\}\label{eq:optim}\\ 
    & \Psi_h(\hat{T_{r}^m},T_{s}) \leq 0, \quad h \in \{1,...,H\}
\end{align*}
where $K$ and $H$ are the number of equality and inequality constraints, respectively. To enable the model to learn and adhere to the specified constraints, we adopt the augmented Lagrangian method, inspired by~\cite{djeumou2022neural}. 
This method involves introducing penalty terms into the objective function to account for constraint violations. 
To do so,  
we further convert the constraints to differentiable terms. 
For instance, we leverage hyperbolic functions to represent a smoothed step function. 

For each constraint in (\ref{eq:optim}), we define a separate Lagrange variable. We define a variable $\lambda^{eq}_{k,i}$ for each equality constraint $\Phi_k$ evaluated at each training data $(\hat{T_{r_i}},T_{r_i})$, and similarly $\lambda^{ineq}_{h,i}$ at each point $(\hat{T_{r_i}},T_{r_i})$ for all the inequality constraints $\Psi_h$. 
The augmented Lagrangian loss function is then given by:
\begin{align*}
\begin{split}
    {L_{aug}} = {}& {L_{combine}(\hat{T_{r}^m},T_{r})} + \sum_{\substack{i\in N \\ k \in K}}\mu\Phi_k(\hat{T_{r_i}^m},T_{s_i})^2 \\
    &+ \sum_{\substack{i\in N \\ k \in K}}\lambda^{eq}_{k,i}\Phi_k(\hat{T_{r_i}^m},T_{s_i}) + \sum_{\substack{i\in N \\ h \in H}}\lambda^{ineq}_{h,i}\Psi_h(\hat{T_{r_i}^m},T_{s_i},)\\
    & + \sum_{\substack{i\in N \\ h \in H}}\mu[\lambda^{ineq}_{h,i}>0\lor\Psi_h>0]\Psi_h(\hat{T_{r_i}^m},T_{s_i})^2 \\
\end{split}
\end{align*}
where $N$ is the training dataset size and $\mu$ is penalty coefficient.
We initialize $\mu$ to be $1\mathrm{e}{-3}$ and $\lambda$ to be 0.
During each iteration of training, we minimize $L_{aug}$ via gradient descent while keeping the values of $\mu$ and $\lambda$ fixed. 
After the transformer model has converged, we update $\mu$ and $\lambda$ according to the update rules:
\begin{align*}
    & \mu \leftarrow \mu * \mu_{mult} \\
    & \lambda^{eq}_{k,i} \leftarrow \lambda^{eq}_{k,i} + 2 * \mu * \Phi_k(\hat{T_{r_i}},T_{s_i}) \\
    & \lambda^{ineq}_{h,i} \leftarrow (\lambda^{ineq}_{h,i} + 2 * \mu * \Psi_h(\hat{T_{r_i}},T_{s_i}))_{+} \\
    & where  \;  x_{+} = max\{0, \; x\} 
\end{align*}
where $\mu_{mult}$ is a hyperparameter of value 1.5.
Then we start another round of training on the transformer model with the updated Lagrange variables. This process is repeated until the constraint violations reach a saturation point and stop decreasing.
In each iteration, the Lagrange multipliers $\lambda$ are updated by incrementing them based on the violations of the corresponding output data multiplied by $\mu$. 
The importance of a violation in the loss function increases as the violation magnitude becomes higher, requiring more effective minimization. 
Training with these penalty terms enables the model to learn the consistency between the input and output, enforced by the constraints.

\subsection{Constraint Enforcement Module (CEM)}
\label{sec:cem}

While the incorporation of constraints in the loss function improves the imputation accuracy, it still provides no guarantee that the constraints will be satisfied. Thus, we introduce the  Constraint Enforcement Module (CEM) which aims at correcting the output of the transformer (\ie forces it to satisfy the specified constraints) while changing it as little as possible.
CEM uses the SMT solver Z3~\cite{z3} to correct the output of the ML model according to the constraints (\ref{eq:constr_eqal},~\ref{eq:constr_ineqal}).
We use variables $\hat{T}_{r}[t]$ to denote the corrected output at each time step. 
To ensure that the corrected time series remains close to the ML model's output, we use the following objective that minimizes the total difference between the corrected and original values, ignoring the time steps in which the data is sampled.
\begin{equation*}
    \min \sum_{t=0, \ t \notin T_{samples}}^{T-1} \vert \hat{T}_{r} [t] - \hat{T}_{r}^m [t] \vert
\end{equation*}

\subsection{Target refinement}
\label{sec:refine}
\begin{figure}[t!]
    \centering
    \includegraphics[width=\linewidth]{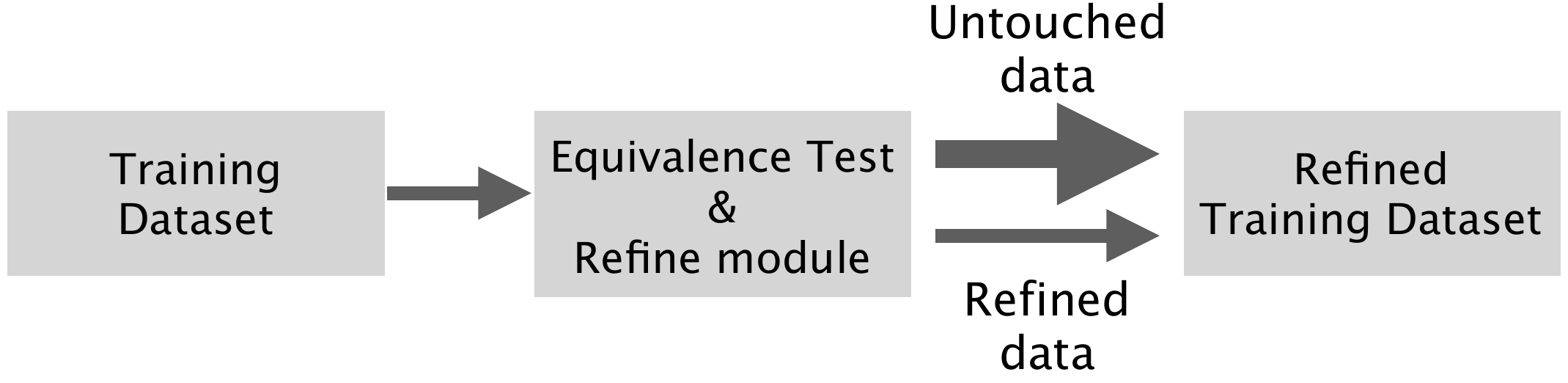}
    \captionsetup{skip=3pt}
    \caption{Training dataset goes through an equivalence test where a small portion of equivalent data is consolidated into one class while the majority remains untouched, forming a refined training dataset.\vspace{-1em}}
    \label{fig:refine}
\end{figure}

As we discussed in~\S\ref{sec:challenge}, certain scenarios involve the same coarse-grained input \coarse occurring multiple times in the training dataset, each time associated with a distinct fine-grained target. 
This one-to-multiple mapping poses challenges for the training convergence of the transformer and risks the usefulness of the result. 
To address this challenge, we designed a target refinement module shown in Fig~\ref{fig:refine}. The module acts solely on training data and is composed of an equivalence test and a refinement mechanism. We first discuss how to refine the distinct targets for the same coarse-grained input, and then delve into how to identify the training data that needs to be refined. 

First, we observe that although the distinct target values associated with the same input exhibit variations in values,
they share the same trends and patterns.
If presented with only a single target,
the transformer model should learn these shared patterns effectively.
Therefore, we aim to provide the model with the same set of plausible targets for the same input in the training set.
We use a set of targets instead of one because all target values are valid and equally plausible.
To achieve this, we consolidate the different targets associated with the same input \coarse into a class ${T_{r}^{class}}$.
We define the loss $L_{class}$ to be the minimum difference between the transformer output $\hat{T_{r}^m}$ and each target in the class.
\begin{equation*}
    L_{class} = \min_{\forall T_{r} \in T_{r}^{class}} {L_{combine}(\hat{T_{r}^m}, T_{r})}
\vspace{-0.15in}
\end{equation*}
This encourages the transformer to match its output to the closest target and backpropagate the difference. 
Even when the same input occurs multiple times, the transformer matches the same target and effectively learns the patterns.

A significant challenge within this approach is defining what constitutes a class that encapsulates data with \textit{same} inputs but \textit{different} outputs. 
Inputs are rarely numerically identical, and determining the proximity of inputs that should result in the same output is challenging due to the complexity of transformer models with thousands of parameters. 
To address this challenge, we train a basic transformer using the raw training dataset and $L_{combine}$ as the loss function.
After training the basic transformer,
instances with close imputed outputs indicate that the transformer cannot distinguish the difference in their inputs, implying that the inputs are close enough.
By leveraging the output from this basic transformer, we identify inputs in the training dataset as part of the same group if:
\textit{(i)} their fine-grained targets are far apart;
\textit{(ii)} their basic transformer outputs are close.
This forms our equivalence test.
The target refinement module facilitates the convergence of the transformer to 
learn essential correlations and
output an adequate result for a given coarse-grained input in the training set.

%% file: eval.tex
\section{Evaluation }
We evaluate \sys across different case studies, using synthetic and real-world data, and we compare it with state-of-the-art approaches. 
Our evaluation aims to answer the following key questions on both imputation accuracy and downstream task accuracy:

\vspace*{0.02in}\noindent(Q1) \;How does \sys perform compared to directly using coarse-grained data? \label{q1}

\vspace*{0.02in}\noindent(Q2) \; How does \sys perform against statistical baselines and state-of-the-art time-series imputation models? \label{eq:q2}

\vspace*{0.02in}\noindent(Q3) \; How does the performance of \sys %
change when we increase the zoom-in factor? \label{eq:q3}
\subsection{Methodology}
We compare \sys against statistical and ML methods. 

\myitem{K-nearest neighbors (KNN)}: 
KNN is a straightforward yet effective technique that has been used for image super-resolution tasks.
For a given coarse-grained data, KNN identifies the nearest K training data inputs and calculates the average of the K labels as the output. The choice of K is determined through experimentation to yield optimal performance.

\myitem{IterativeImputer~\cite{scikit-learn}}: This is a statistical method that iteratively models features with missing values 
as linear functions of other features iteratively, retaining the periodic samples. 
To incorporate other measurements such as the maximum value, we place them at the midpoint of each interval. Essentially, the results represent all coarse-grained data.

\myitem{Plain transformer}: This is a transformer model trained using MSE (without our improvements \eg the knowledge incorporation, or the refinement step). 

\myitem{Brits~\cite{cao2018brits}}: Brits employs bidirectional recurrent neural networks for imputing missing values in time series data.
To adapt Brits to our settings, we incorporate sampled values such as sum and max by placing them at the end of the time interval and use Brits to impute values between periodic samples. 

\myitem{Metrics.}
We evaluate \sys and our baselines by their imputation accuracy and their performance in downstream tasks. 
To quantify imputation accuracy, we calculate autocorrelation, distance, and distribution differences between imputed time series and the ground truth.
Next, we evaluate the quality of imputation by comparing the performance of downstream tasks using the imputed results as input against using the ground truth. 
For each of the metrics, we report average $relative\_error = \frac{|t - t_{real}|}{t_{real}}$ over the testing dataset, where $t_{real}$ is the ground truth of a metric and $t$ is the measured value.  
Because the errors have very different scales over different methods, we normalize the relative errors of each metric to [0.1, 0.9] for better visualization.

\myitem{Case studies and goals.} We consider three case studies to show \sys capabilities. For each of these, we first explain an example scenario in which operators have certain downstream tasks in mind. Next, we explain how we use an existing or synthetic dataset to evaluate \sys in this scenario. While the dataset we use contains fine-grained time series, we treat this as ground truth; thus, we assume it is not available to the operator to use directly. This is realistic as it is effectively equivalent to collecting very fine-grained data for a very short period of time to train on. The downstream tests run on input that is calculated by \sys (or by other baselines). The input of \sys (and of the other  baselines) is the coarse-grained version of each time series in the dataset.

\begin{figure*}
\begin{subfigure}{0.5\columnwidth}
  \centering
  \includegraphics[width=1\columnwidth]{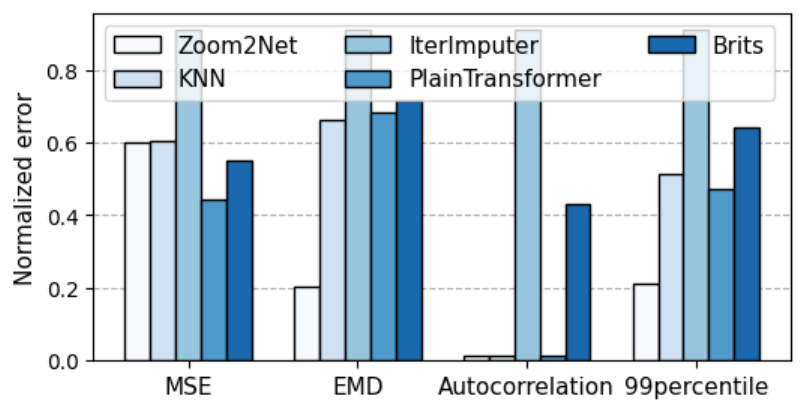}
  \caption{Case-1 Synthetic}
  \label{fig:queue-stats}
\end{subfigure}
\hfill
\begin{subfigure}{0.5\columnwidth}
  \centering
	\includegraphics[width=1\columnwidth]{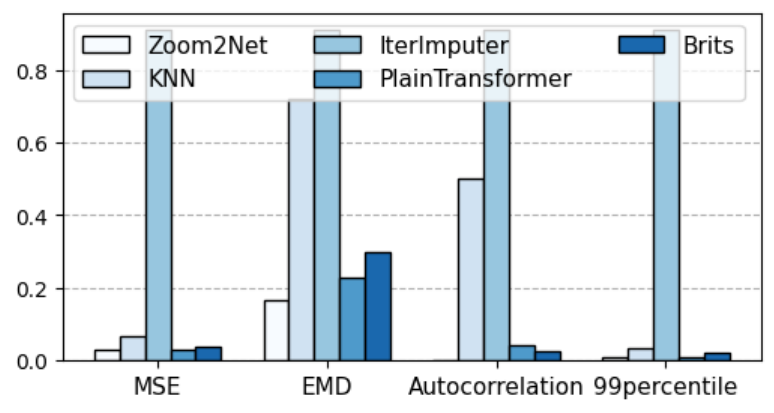}
	\caption{Case-1 Meta}
	\label{fig:meta-stats}
\end{subfigure}
\hfill
\begin{subfigure}{0.5\columnwidth}
  \centering
	\includegraphics[width=1\columnwidth]{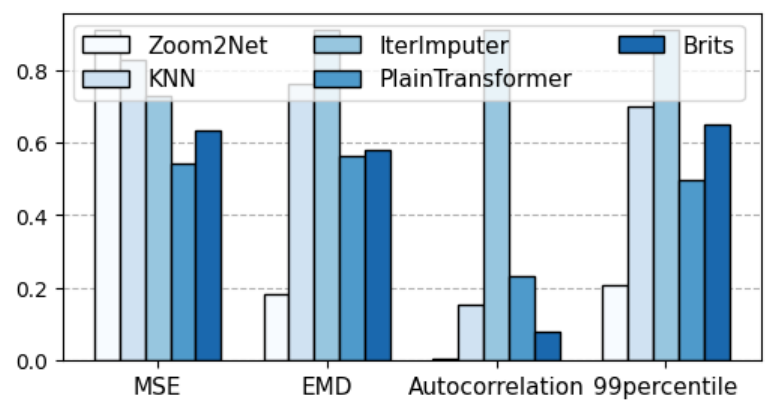}
	\caption{Case-2 MLab}
	\label{fig:mlab-stats}
\end{subfigure}
\hfill
\begin{subfigure}{0.5\columnwidth}
  \centering
	\includegraphics[width=1\columnwidth]{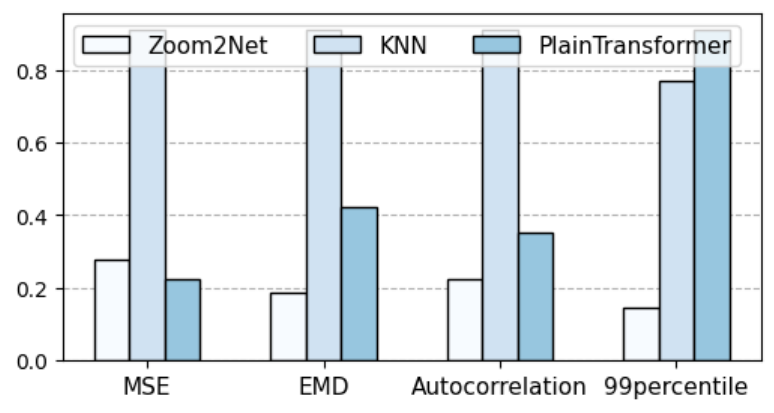}
	\caption{Case-3 VPN}
	\label{fig:vpn-stats}
\end{subfigure}
\captionsetup{skip=3pt}
\caption{\sys outperforms baselines in EMD, Autocorrelation, and 99p accuracy, leading to superior performance in downstream tasks. The plain transformer achieves the lowest MSE by generating overly smooth results.}
\label{fig:onequeue}
\end{figure*}
\subsection{Case study 1: ToR burstiness in a Cloud}
\label{sec:case1}

Consider an operator of 
a large data center who has to run a set of downstream tasks, e.g., deciding how much on-chip buffer to provision to network switches, or detecting adversarial traffic patterns. 
To inform these tasks, the operator needs to learn about burst properties in queue lengths, specifically burst position, height, frequency, inter-arrival distance, duration, and volume. 
Accurate analysis requires fine-grained switch queue length measurements at the millisecond level. 
Alternatively, they can use link utilization measurements as a proxy for queue length collected at data center RTT granularity, \ie 1ms, as demonstrated by researchers at Meta~\cite{ghab2022millisamp}. 
We test both cases using a synthetic dataset and a dataset released by Meta.

\myitem{Synthetic Dataset.}
We generate a dataset using ns-3 simulator~\cite{ns3}, simulating a leaf-spine topology as described in~\cite{ns3abm}. 
The switches in the simulation adhere to the features of Broadcom TridentII~\cite{trident2} and are configured with Dynamic Thresholds~\cite{choudhury1998dynamic} as buffer management scheme.
The generated traffic 
follows web search and incast traffic patterns, incorporating various settings for traffic load, burst size, burst frequency, and congestion control algorithms (\eg DCTCP and Cubic). 
During simulation, we collect fine-grained time series (ground truth) including queue lengths, per-port packet, and drop counts every 1ms. 
We generate coarse-grained time series by sampling the fine-grained ones at 50ms granularity, mimicking the following monitoring tools: \ie \emph{(i)} LANZ~\cite{lanz}, which provides the per-queue maximum length within each interval, \emph{(ii)} SNMP~\cite{snmp1}, which provides per-port counts of packets sent and dropped every interval; and \emph{(iii)} periodic sampling. Our training dataset contains 8,000 data points.
\textbf{Imputation goal: }\sys takes maximum, periodic sampled queue length, packets dropped, and packets sent count at \textit{50ms} granularity and produces \textit{1ms} fine-grained queue lengths.

\myitem{Meta Dataset.}
We use a public dataset from Meta~\cite{ghab2022millisamp} 
which contains 
link utilization, retransmission traffic, in-congestion traffic, and connection counts at a fine-grained resolution of 1ms.
We aggregate them at intervals of 50ms to create coarse-grained data.
In total, we use a training set of 20,000 data points.
\textbf{Imputation goal: }\sys takes the aggregated measurements at \textit{50ms} granularity and produces \textit{1ms} fine-grained link utilization.

\myitem{Synthetic data constraints.}
We use three constraints on imputed queue lengths $\hat{T}_{r}$ for every coarse time interval $T$ (50ms). 

Measurement constraints are simple inversions of the known functions used for coarsening (i.e., the monitoring tools). Concretely, we require that the maximum value of the imputed queue length time series at every interval equals the value that LANZ reported $m\_max$, and the instantaneous queue length $m\_len_{t}$ equals the value of the periodic sampling at $t^{th}$ms. 

\begin{align}
\tag{$C1$}
   &\max_{0 \leq t < T} \hat{T}_{r} [t] = m\_max\label{eq:max}\\
   \tag{$C2$}
   &\forall t \in T_{samples}.\quad \hat{T}_{r} [t] = m\_len_{t} 
   \label{eq:sampling}
\end{align}
Operational constraints 
express the connection between switch operation and the counts of packets sent from SNMP measurements.
If a queue is nonempty for $NE$ ms, then at least $NE$ packets have been dequeued, as 
schedulers are work-conserving.
An empty queue can send a packet if one arrives; hence $NE$ is a lower bound on packets sent count ($m\_out$).
\begin{align*}
\tag{C3}
    NE &\leq m\_out \label{eq:num_sent}\\
 where \quad
 NE &= \sum_{t = 0}^{T-1} ite( \hat{T}_{r}[t] > 0,\quad 1,\quad 0)
\end{align*}
\myitem{Meta data constraints.}
We formulate four constraints for imputed link utilization $\hat{T}_{r}$ for every coarse time interval $T$ (\ie 50ms). Measurement constraints come from the measurement of aggregated traffic rates $m\_sum$.
\begin{align}
\tag{$C4$}
\sum_{t = 0}^{T-1} \hat{T}_{r} [t] = m\_sum \label{eq:sum}
\end{align}

Operational constraints for imputed link utilization data articulate its relationships with congestion and retransmission. The imputed link utilization should be at least the number of bytes in both congestion ($congestion\_sum$) and retransmitted ($retransmit\_sum$) scenarios.
In the presence of congestion during a 50ms interval, there should be at least one burst observed in the imputed link utilization. 
\begin{align*}
   \tag{$C5$}
   \label{eq:retrans}
   \sum_{t = 0}^{T-1} \hat{T}_{r} [t] &\geq retransmit\_sum \\
   \tag{$C6$}
   \sum_{t = 0}^{T-1} \hat{T}_{r} [t] &\geq congestion\_sum \\
   \tag{C7}
    congestion\_sum > 0  &\rightarrow \max_{0 \leq t < T} \hat{T}_{r} [t] \geq \frac{1}{2}bandwidth
\end{align*}

\begin{figure*}
\begin{subfigure}{0.33\textwidth}
  \centering
  \includegraphics[width=\columnwidth]{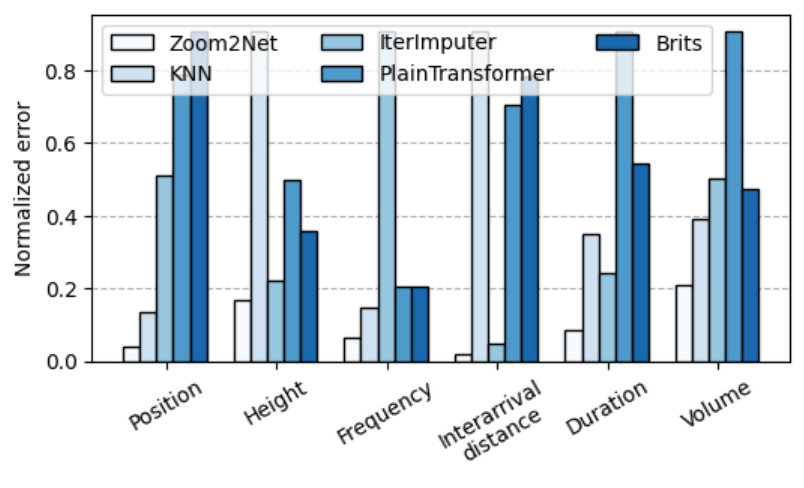}
  \captionsetup{skip=3pt}
  \caption{Burst analysis for Case-1 synthetic}
  \label{fig:queue-downstream}
\end{subfigure}
\begin{subfigure}{0.33\textwidth}
  \centering
	\includegraphics[width=\columnwidth]{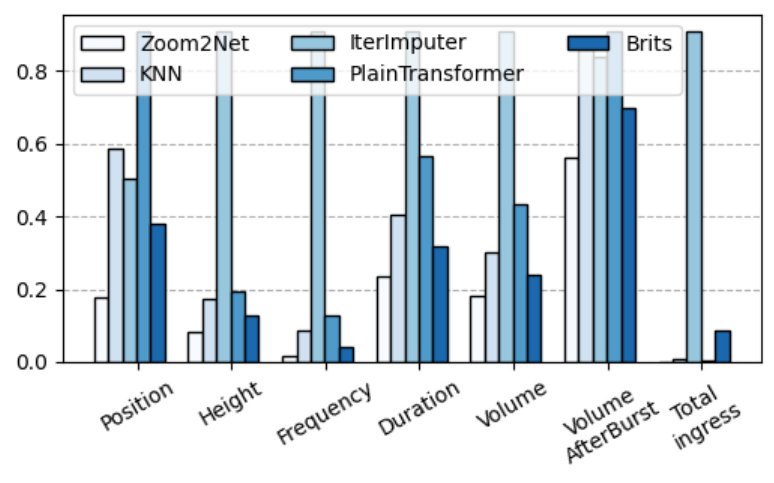}
 \captionsetup{skip=3pt}
	\caption{Burst analysis for Case-1 Meta}
	\label{fig:meta-downstream}
\end{subfigure}
\begin{subfigure}{0.33\textwidth}
  \centering
	\includegraphics[width=\columnwidth]{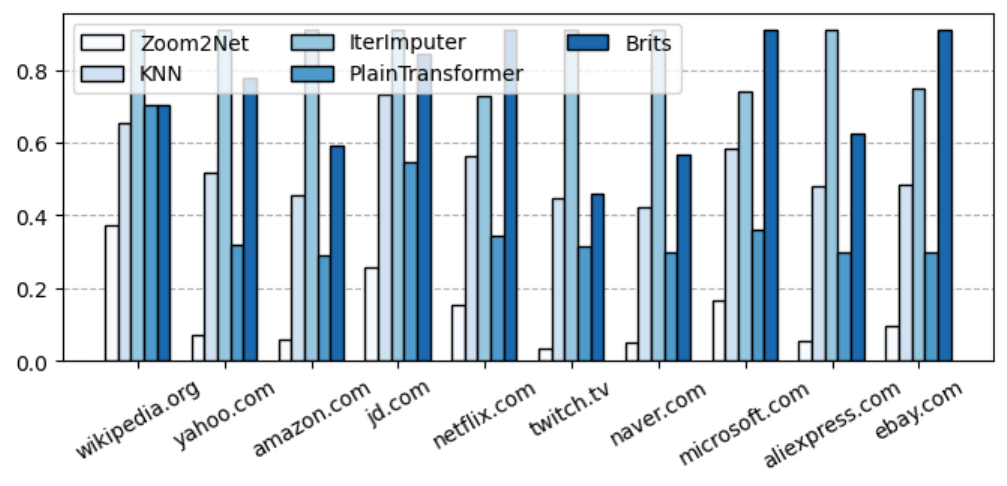}
 \captionsetup{skip=3pt}
	\caption{Page loading time for Case-2 MLab}
	\label{fig:mlab-downstream}
\end{subfigure}
\captionsetup{skip=3pt}
\caption{Normalized error for downstream task performance. (a) and (b) show that \sys effectively captures burst behaviors in both synthetic and real-world datasets. (c) demonstrates that \sys precisely estimates webpage loading time for various websites by imputing accurate sending rates.
Overall, \sys surpasses baselines by 38\%.}
\end{figure*}

\myitem{\sys captures the structural pattern of the fine-grained time series better than baselines, although it results in higher MSE (Q2).} 
Fig.~\ref{fig:queue-stats} and Fig.~\ref{fig:meta-stats} summarize the metrics for \sys accuracy on both datasets. 
\sys outperforms baselines in EMD, autocorrelation, and the 99th percentile by 33--53\% in the synthetic dataset.
Unsurprisingly, the plain transformer, which is trained with MSE, has the lowest error in MSE, as it generates overly smooth and, hence, low values for queue lengths.
On the contrary, the incorporation of EMD in the loss function prompts \sys to de-emphasize minor shifts in burst positions, which hurts MSE but encourages more accurate imputation of the burst shape. 
For Meta data, the three metric accuracies (EMD, autocorrelation, and the 99th percentile) of \sys imputed link utilization is better than baselines by a margin of 33\%.

\myitem{\sys effectively recovers bursts, outperforming other baselines in all downstream tasks (Q2).}
As shown in Fig.~\ref{fig:queue-downstream}, \sys achieves a significant improvement in burst properties ranging from 10\% to 88\% over all baselines in all tasks for synthetic data.
Notably, \sys attains an average error of only 4\% for burst position, even though \sys loss (including EMD) treats slightly shifted bursts as equivalent, demonstrating its effectiveness in capturing network dynamism. 
Across burst analysis for the real-world data in Fig.~\ref{fig:meta-downstream}, \sys exhibits an average performance superiority of 30\%.
There are cases where certain baselines perform comparably with \sys, such as plain transformer and Brits.
This can be attributed to 
the characteristics of the dataset.
The data is skewed towards smaller values with the infrequent occurrence of bursts lasting 1-2ms, leading to low errors in the 99th percentile, autocorrelation, 
burst height and frequency.

\myitem{\sys enhances performance compared to directly using coarse-grained data (Q1).}  \sys achieves up to 5 times better performance in downstream tasks compared to the IterImputer, which embeds both the maximum queue length and periodic samples. IterImputer can be seen as a way to use coarse-grained data directly, as it does not incorporate any learning.
This highlights the critical role of \sys in learning from data to improve downstream tasks.

\myitem{\sys leverages correlations learned in different settings.}
The synthetic testing dataset 
includes combinations of traffic patterns and congestion control algorithm settings that were not present in the training set. 
Remarkably, \sys performs even better on these unseen settings by an average of 30\%
compared to the scenarios in the training set.
This underscores \sys's capacity to apply learned correlations effectively to diverse, previously unseen scenarios.

\subsection{Case study 2: CDN PoP selection}\label{sec:case3}
Consider a Content Delivery Network (CDN) operator managing multiple Point-of-Presence (PoP) and serving different webpages to users. 
When determining the optimal PoP for serving diverse webpages to diverse users, the operator needs to estimate the time it takes for different amounts of data (corresponding to each webpage) to reach users from each PoP.  Observe that because of congestion control and the heterogeneity of networks, predicting time to transfer is not trivial ,i.e., it is not a linear connection.
The operator can calculate the time required to send varying amounts of webpage data from each PoP to each user around the globe by observing the sending rate of the user-PoP pair over time. In practice, the operator cannot send varying amounts of data to users and record the time of reach. Instead, they have access to user-initiated network speed tests using Network Diagnostic Tools (NDT) at a coarser granularity.

\myitem{MLab Datasets.}
For this demonstration, we leverage data from the M-Lab project's NDT measurements~\cite{phillipa2022mlab}.
The NDT measurements capture TCPInfo and BBRInfo statistics from each snapshot of a data transfer that spans 250ms on average. These statistics include achieved throughput, minimum RTT, bytes sent, retransmitted, and more.
M-Lab also provides packet traces recorded during NDT measurements. These traces provide transmission rates, forming the fine-grained time series at 10ms granularity. Our training set contains 5000 data points. 
\textbf{Imputation goal: }\sys leverages the coarse-grained NDT measurements recorded at \textit{250ms} intervals, to impute fine-grained sending rates at a finer granular level of \textit{10ms}.

\myitem{Constraints.}
We formulate two measurement constraints related to the maximum sending rate and aggregated sending traffic volume, expressed similarly as (\ref{eq:max}) and  (\ref{eq:sum}).

For operational constraints, the first one ensures that traffic volume is at least the number of bytes retransmitted, similar to (\ref{eq:retrans}).
Furthermore, when the NDT measurement period ($elapsed\_time$) is shorter than the RTT, the sending traffic should not exceed the product of the Maximum Segment Size ($MSS$) and congestion window size ($SndCwnd$). 
If the measurement period is shorter than the time spent waiting for the receiver window ($RwndLimited$), indicating the sender waiting and pauses sending, the traffic rate should remain at 0.
\begin{align*}
\tag{$C8$}
   & elapsed\_time \leq RTT   \rightarrow \sum_{t = 0}^{T-1} \hat{T}_{r} [t] \leq MSS \times SndCwnd\\
   \tag{$C9$}
   &elapsed\_time \leq RwndLimited \rightarrow  \sum_{t = 0}^{T-1} \hat{T}_{r} [t] =0 
\end{align*}

\myitem{\sys's fine-grained output provides an accurate estimate of page loading time to users around the globe.}
Fig.~\ref{fig:mlab-downstream} illustrates the average error of page loading time estimation for the top 10 Alexa websites. Page loading time is emulated as the time it takes to transfer the same amount of data as the website to the client.
The number of bytes transmitted by these websites varies from 300KB to 10MB. 
\sys shows an average improvement of 43\% in accuracy compared to other baselines.
In Fig.~\ref{fig:mlab-stats}, \sys shows better performance by 44\% on average across statistics metrics.

\subsection{Case study 3: Encrypted traffic \\ classification}\label{sec:case4}
Consider a network practitioner tasked with detecting encrypted VPN traffic based on
flow-based time-related features, such as flow duration, maximum, minimum, and average forward/backward inter-arrival time~\cite{draper2016characterization}.
The practitioner had access to a packet trace, which they used to extract these features, train their model, and then discard the trace due to its impractically large size for storage and privacy concerns.
Some months later,  the practitioner discovers new features useful for classification, but they cannot retrain because the original traces are no longer available for additional feature extraction.
\sys offers a solution in this scenario. When the practitioner initially extracts features from the trace, they can use these features as coarse-grained input for \sys to impute the packet trace, aiming to recover details such as the arrival time and length of each packet. 
The practitioner can then keep the \sys model and discard the trace.
Later, they can extract new features from the imputed trace
 and add to classification tasks.
This provides the flexibility to extract new features without the overhead of storing large packet traces,
enabling a more adaptive approach to feature engineering and classification. 

\myitem{VPN Datasets.}
In this case study, we begin with real-world packet trace data~\cite{draper2016characterization}. From this data, we extract features such as flow duration, maximum, minimum, and average forward/backward inter-arrival time, and maximum packet length which serve as the coarse-grained input.
We formulate the fine-grained data, specifically packet arrival time and packet length by parsing the trace. We use a training set of 3,300 data points.
\textbf{Imputation goal: }\sys uses a \textit{single} set of coarse-grained features extracted from packet traces to generate fine-grained packet trace information of averaged \textit{20} packets.
From imputed packet trace information, we extract additional features flow rates and add them to the initial features for classification.

\myitem{Constraints.}
We utilize measurement constraints specific to the mean, minimum and maximum of forward/backward inter-arrival time, and maximum of packet length. They are formulated similar to (\ref{eq:max}).
\begin{figure}[t!]
        \centering
\includegraphics[width=0.9\linewidth]{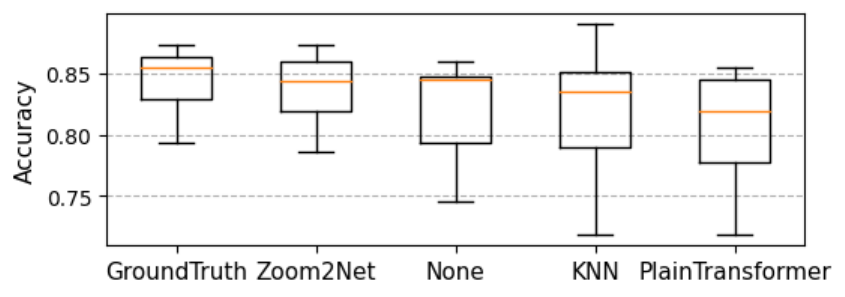}
    \captionsetup{skip=3pt}
    \caption{\sys improves accuracy and stability of VPN traffic classification by recovering features from imputed traces.}
    \label{fig:vpn-downstream}
    \vspace{-1em}
\end{figure}

\myitem{\sys can recover features by imputing packet traces, improving traffic classification accuracy.} 
In Fig.~\ref{fig:vpn-downstream},
we report the classification accuracy of a multi-layer perceptron model that uses the union of ground-truth features calculated on the initial trace with additional features calculated on the imputed trace. 
The different bars illustrate the 
source of the additional features:
ground-truth traces, KNN-imputed traces, plain transformer-imputed traces, and the case where no additional feature is added. Each scenario runs for 10 times.
We observe that the classification accuracy using features extracted from \sys-imputed traces is comparable to that extracted from the ground truth and
higher than scenarios without additional features, demonstrating the effectiveness of \sys in capturing the correlations and characteristics of traces. 
Notably, the classifier trained with \sys-imputed features is more stable.
The minimum, the first quartile, and the third quartile of Zoom2Net's accuracies are the highest among baselines. 
For statistical metrics in Fig.~\ref{fig:vpn-stats}, \sys demonstrates a notable average performance improvement of 53\% over other methods.
The given dataset lacks periodic samples, restricting the applicability of IterImputer and Brits so we did not compare with them.

\subsection{Zoom-in factor analysis}
\begin{figure}[t]
     \centering
     \begin{subfigure}[b]{0.48\columnwidth}
         \centering
         \includegraphics[width=1\textwidth]{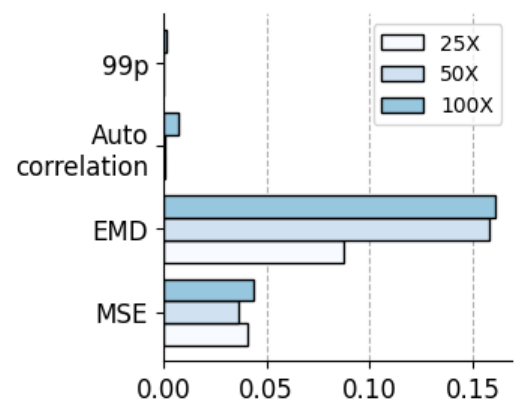}
         \caption{Normalized relative error of imputation accuracy}
         \label{fig:zoom_stats}
     \end{subfigure}
     \hfill
     \begin{subfigure}[b]{0.48\columnwidth}
         \centering
         \includegraphics[width=1\textwidth]{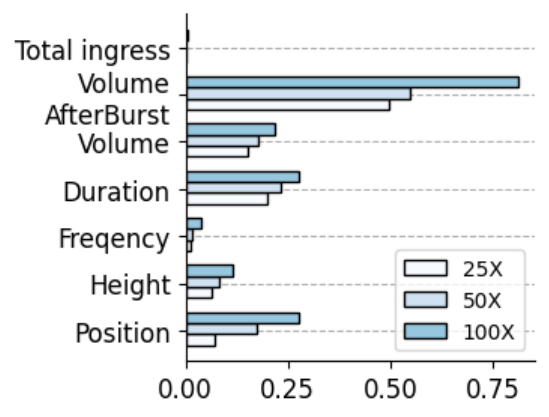}
         \caption{Normalized relative error of downstream task accuracy}
         \label{fig:zoom_downstream}
     \end{subfigure}
     \caption{\sys does not exhibit substantial performance degradation under different zoom-in factors.}
\label{fig_temporal_trace}
\vspace{-0.15in}
\end{figure}

In this section, we evaluate the performance of \sys under different zoom-in factors, defined as the ratio of coarse granularity and fine granularity. 
Using the Meta dataset from~\S\ref{sec:case1}\\~\cite{ghab2022millisamp},
we evaluate three zoom-in factors: 25, 50, and 100. We calculate their relative error compared to the ground truth and normalize the error using the same normalization factor as in~\S\ref{sec:case1} to facilitate comparisons with other baseline methods in terms of structural metrics and downstream tasks.
Fig.~\ref{fig:zoom_stats} shows that imputation accuracy does not significantly vary as we increase the zoom-in factor. Remarkably, the imputation accuracy with a factor of 100 outperforms baselines with a factor of 50 by an average of 32\%.
Fig.~\ref{fig:zoom_downstream} shows that the downstream task accuracy of the three zoom-in factors does not exhibit substantial differences except for the volume-after-burst metric. Nevertheless, the overall performance remains superior to baselines by 23\% on average. 
These results underscore \sys's capability to effectively impute super coarse-grained data to fine-grained measurements.

%% file: conclusion.tex
\section{Conclusion}
This paper presents a new paradigm for network telemetry. Instead of improving hardware or collection algorithms, we advocate for post-collection software telemetry imputation. We present \sys, an ML-based system that analyzes multiple correlated coarse-grained time series to impute their fine-grained counterparts. What sets \sys apart is its incorporation of domain knowledge through operational and measurement constraints. 
We explored several use cases for \sys
using synthetic and public datasets and demonstrated \sys's capability to accurately impute diverse types of telemetry data. The results highlight the effectiveness of \sys in facilitating reliable downstream tasks.

\newpage

%% file: paper.bbl

\begin{thebibliography}{45}


\ifx \showCODEN    \undefined \def \showCODEN     #1{\unskip}     \fi
\ifx \showDOI      \undefined \def \showDOI       #1{#1}\fi
\ifx \showISBNx    \undefined \def \showISBNx     #1{\unskip}     \fi
\ifx \showISBNxiii \undefined \def \showISBNxiii  #1{\unskip}     \fi
\ifx \showISSN     \undefined \def \showISSN      #1{\unskip}     \fi
\ifx \showLCCN     \undefined \def \showLCCN      #1{\unskip}     \fi
\ifx \shownote     \undefined \def \shownote      #1{#1}          \fi
\ifx \showarticletitle \undefined \def \showarticletitle #1{#1}   \fi
\ifx \showURL      \undefined \def \showURL       {\relax}        \fi
\providecommand\bibfield[2]{#2}
\providecommand\bibinfo[2]{#2}
\providecommand\natexlab[1]{#1}
\providecommand\showeprint[2][]{arXiv:#2}

\bibitem[\protect\citeauthoryear{??}{ns3}{2023}]%
        {ns3abm}
 \bibinfo{year}{2023}\natexlab{}.
\newblock \showarticletitle{{ns3-datacenter}}. \bibinfo{howpublished}{\url{https://github.com/inet-tub/ns3-datacenter}}.
\newblock


\bibitem[\protect\citeauthoryear{Apostolaki, Singla, and Vanbever}{Apostolaki et~al\mbox{.}}{2021}]%
        {RouteScout}
\bibfield{author}{\bibinfo{person}{Maria Apostolaki}, \bibinfo{person}{Ankit Singla}, {and} \bibinfo{person}{Laurent Vanbever}.} \bibinfo{year}{2021}\natexlab{}.
\newblock \showarticletitle{{Performance-Driven Internet Path Selection}}. In \bibinfo{booktitle}{{\em Proceedings of the ACM SIGCOMM Symposium on SDN Research (SOSR)}}. \bibinfo{pages}{41--53}.
\newblock


\bibitem[\protect\citeauthoryear{Arista}{Arista}{2016}]%
        {lanz}
\bibfield{author}{\bibinfo{person}{Arista}.} \bibinfo{year}{2016}\natexlab{}.
\newblock \showarticletitle{Arista LANZ Overview}. \bibinfo{howpublished}{\url{https://arista.com/assets/data/pdf/Whitepapers/Arista_LANZ_Overview_TechBulletin_0213.pdf}}.
\newblock


\bibitem[\protect\citeauthoryear{Bahat and Michaeli}{Bahat and Michaeli}{2020}]%
        {bahat2020explorable}
\bibfield{author}{\bibinfo{person}{Yuval Bahat} {and} \bibinfo{person}{Tomer Michaeli}.} \bibinfo{year}{2020}\natexlab{}.
\newblock \showarticletitle{{Explorable Super Resolution}}. In \bibinfo{booktitle}{{\em Proceedings of the IEEE/CVF Conference on Computer Vision and Pattern Recognition (CVPR)}}.
\newblock


\bibitem[\protect\citeauthoryear{Ben~Basat, Ramanathan, Li, Antichi, Yu, and Mitzenmacher}{Ben~Basat et~al\mbox{.}}{2020}]%
        {ben2020pint}
\bibfield{author}{\bibinfo{person}{Ran Ben~Basat}, \bibinfo{person}{Sivaramakrishnan Ramanathan}, \bibinfo{person}{Yuliang Li}, \bibinfo{person}{Gianni Antichi}, \bibinfo{person}{Minian Yu}, {and} \bibinfo{person}{Michael Mitzenmacher}.} \bibinfo{year}{2020}\natexlab{}.
\newblock \showarticletitle{PINT: Probabilistic In-band Network Telemetry} {\em (\bibinfo{series}{SIGCOMM '20})}. \bibinfo{publisher}{Association for Computing Machinery}, \bibinfo{address}{New York, NY, USA}.
\newblock
\showISBNx{9781450379557}
\showDOI{%
\url{https://doi.org/10.1145/3387514.3405894}}


\bibitem[\protect\citeauthoryear{Broadcom}{Broadcom}{2023}]%
        {trident2}
\bibfield{author}{\bibinfo{person}{Broadcom}.} \bibinfo{year}{2023}\natexlab{}.
\newblock \showarticletitle{{Trident2 / BCM56850 Series, High-Capacity StrataXGS® Trident II Ethernet Switch Series}}. \bibinfo{howpublished}{\url{https://www.broadcom.com/products/ethernet-connectivity/switching/strataxgs/bcm56850-series}}.
\newblock


\bibitem[\protect\citeauthoryear{Brown, Mann, Ryder, Subbiah, Kaplan, Dhariwal, Neelakantan, Shyam, Sastry, Askell, Agarwal, Herbert-Voss, Krueger, Henighan, Child, Ramesh, Ziegler, Wu, Winter, Hesse, Chen, Sigler, Litwin, Gray, Chess, Clark, Berner, McCandlish, Radford, Sutskever, and Amodei}{Brown et~al\mbox{.}}{2020}]%
        {brown2020gpt}
\bibfield{author}{\bibinfo{person}{Tom Brown}, \bibinfo{person}{Benjamin Mann}, \bibinfo{person}{Nick Ryder}, \bibinfo{person}{Melanie Subbiah}, \bibinfo{person}{Jared~D Kaplan}, \bibinfo{person}{Prafulla Dhariwal}, \bibinfo{person}{Arvind Neelakantan}, \bibinfo{person}{Pranav Shyam}, \bibinfo{person}{Girish Sastry}, \bibinfo{person}{Amanda Askell}, \bibinfo{person}{Sandhini Agarwal}, \bibinfo{person}{Ariel Herbert-Voss}, \bibinfo{person}{Gretchen Krueger}, \bibinfo{person}{Tom Henighan}, \bibinfo{person}{Rewon Child}, \bibinfo{person}{Aditya Ramesh}, \bibinfo{person}{Daniel Ziegler}, \bibinfo{person}{Jeffrey Wu}, \bibinfo{person}{Clemens Winter}, \bibinfo{person}{Chris Hesse}, \bibinfo{person}{Mark Chen}, \bibinfo{person}{Eric Sigler}, \bibinfo{person}{Mateusz Litwin}, \bibinfo{person}{Scott Gray}, \bibinfo{person}{Benjamin Chess}, \bibinfo{person}{Jack Clark}, \bibinfo{person}{Christopher Berner}, \bibinfo{person}{Sam McCandlish}, \bibinfo{person}{Alec Radford}, \bibinfo{person}{Ilya Sutskever}, {and}
  \bibinfo{person}{Dario Amodei}.} \bibinfo{year}{2020}\natexlab{}.
\newblock \showarticletitle{Language Models are Few-Shot Learners}. In \bibinfo{booktitle}{{\em Advances in Neural Information Processing Systems}}, \bibfield{editor}{\bibinfo{person}{H.~Larochelle}, \bibinfo{person}{M.~Ranzato}, \bibinfo{person}{R.~Hadsell}, \bibinfo{person}{M.F. Balcan}, {and} \bibinfo{person}{H.~Lin}} (Eds.), Vol.~\bibinfo{volume}{33}. \bibinfo{publisher}{Curran Associates, Inc.}, \bibinfo{pages}{1877--1901}.
\newblock
\showURL{%
\url{https://proceedings.neurips.cc/paper_files/paper/2020/file/1457c0d6bfcb4967418bfb8ac142f64a-Paper.pdf}}


\bibitem[\protect\citeauthoryear{Buehler, Horvath, Lyons, Arribas, and Wood}{Buehler et~al\mbox{.}}{2020}]%
        {buehler2020data}
\bibfield{author}{\bibinfo{person}{Hans Buehler}, \bibinfo{person}{Blanka Horvath}, \bibinfo{person}{Terry Lyons}, \bibinfo{person}{Imanol~Perez Arribas}, {and} \bibinfo{person}{Ben Wood}.} \bibinfo{year}{2020}\natexlab{}.
\newblock \showarticletitle{A data-driven market simulator for small data environments}.
\newblock \bibinfo{journal}{{\em arXiv preprint arXiv:2006.14498\/}} (\bibinfo{year}{2020}).
\newblock


\bibitem[\protect\citeauthoryear{Cao, Wang, Li, Zhou, Li, and Li}{Cao et~al\mbox{.}}{2018}]%
        {cao2018brits}
\bibfield{author}{\bibinfo{person}{Wei Cao}, \bibinfo{person}{Dong Wang}, \bibinfo{person}{Jian Li}, \bibinfo{person}{Hao Zhou}, \bibinfo{person}{Lei Li}, {and} \bibinfo{person}{Yitan Li}.} \bibinfo{year}{2018}\natexlab{}.
\newblock \showarticletitle{BRITS: Bidirectional Recurrent Imputation for Time Series}. In \bibinfo{booktitle}{{\em Advances in Neural Information Processing Systems}}, \bibfield{editor}{\bibinfo{person}{S.~Bengio}, \bibinfo{person}{H.~Wallach}, \bibinfo{person}{H.~Larochelle}, \bibinfo{person}{K.~Grauman}, \bibinfo{person}{N.~Cesa-Bianchi}, {and} \bibinfo{person}{R.~Garnett}} (Eds.), Vol.~\bibinfo{volume}{31}. \bibinfo{publisher}{Curran Associates, Inc.}
\newblock
\showURL{%
\url{https://proceedings.neurips.cc/paper_files/paper/2018/file/734e6bfcd358e25ac1db0a4241b95651-Paper.pdf}}


\bibitem[\protect\citeauthoryear{Chen, Feibish, Koral, Rexford, Rottenstreich, Monetti, and Wang}{Chen et~al\mbox{.}}{2019}]%
        {chen2019conquest}
\bibfield{author}{\bibinfo{person}{Xiaoqi Chen}, \bibinfo{person}{Shir~Landau Feibish}, \bibinfo{person}{Yaron Koral}, \bibinfo{person}{Jennifer Rexford}, \bibinfo{person}{Ori Rottenstreich}, \bibinfo{person}{Steven~A Monetti}, {and} \bibinfo{person}{Tzuu-Yi Wang}.} \bibinfo{year}{2019}\natexlab{}.
\newblock \showarticletitle{Fine-grained queue measurement in the data plane}. In \bibinfo{booktitle}{{\em Proceedings of the 15th International Conference on Emerging Networking Experiments And Technologies}} {\em (\bibinfo{series}{CoNEXT '19})}. \bibinfo{publisher}{Association for Computing Machinery}, \bibinfo{address}{New York, NY, USA}, \bibinfo{pages}{15–29}.
\newblock
\showISBNx{9781450369985}
\showDOI{%
\url{https://doi.org/10.1145/3359989.3365408}}


\bibitem[\protect\citeauthoryear{Choudhury and Hahne}{Choudhury and Hahne}{1998}]%
        {choudhury1998dynamic}
\bibfield{author}{\bibinfo{person}{Abhijit~K Choudhury} {and} \bibinfo{person}{Ellen~L Hahne}.} \bibinfo{year}{1998}\natexlab{}.
\newblock \showarticletitle{Dynamic queue length thresholds for shared-memory packet switches}.
\newblock \bibinfo{journal}{{\em IEEE/ACM Transactions On Networking\/}} \bibinfo{volume}{6}, \bibinfo{number}{2} (\bibinfo{year}{1998}), \bibinfo{pages}{130--140}.
\newblock


\bibitem[\protect\citeauthoryear{De~Moura and Bj{\o}rner}{De~Moura and Bj{\o}rner}{2008}]%
        {z3}
\bibfield{author}{\bibinfo{person}{Leonardo De~Moura} {and} \bibinfo{person}{Nikolaj Bj{\o}rner}.} \bibinfo{year}{2008}\natexlab{}.
\newblock \showarticletitle{{Z3: An efficient SMT solver}}. In \bibinfo{booktitle}{{\em Tools and Algorithms for the Construction and Analysis of Systems: 14th International Conference, TACAS 2008, Held as Part of the Joint European Conferences on Theory and Practice of Software, ETAPS 2008, Budapest, Hungary, March 29-April 6, 2008. Proceedings 14}}. Springer, \bibinfo{pages}{337--340}.
\newblock


\bibitem[\protect\citeauthoryear{Dietm\"{u}ller, Ray, Jacob, and Vanbever}{Dietm\"{u}ller et~al\mbox{.}}{2022}]%
        {dietmuller2022transformer}
\bibfield{author}{\bibinfo{person}{Alexander Dietm\"{u}ller}, \bibinfo{person}{Siddhant Ray}, \bibinfo{person}{Romain Jacob}, {and} \bibinfo{person}{Laurent Vanbever}.} \bibinfo{year}{2022}\natexlab{}.
\newblock \showarticletitle{{A New Hope for Network Model Generalization}}. In \bibinfo{booktitle}{{\em Proceedings of the 21st ACM Workshop on Hot Topics in Networks}}. \bibinfo{pages}{152–159}.
\newblock


\bibitem[\protect\citeauthoryear{Djeumou, Neary, Goubault, Putot, and Topcu}{Djeumou et~al\mbox{.}}{2022}]%
        {djeumou2022neural}
\bibfield{author}{\bibinfo{person}{Franck Djeumou}, \bibinfo{person}{Cyrus Neary}, \bibinfo{person}{Eric Goubault}, \bibinfo{person}{Sylvie Putot}, {and} \bibinfo{person}{Ufuk Topcu}.} \bibinfo{year}{2022}\natexlab{}.
\newblock \showarticletitle{Neural Networks with Physics-Informed Architectures and Constraints for Dynamical Systems Modeling}. In \bibinfo{booktitle}{{\em Proceedings of The 4th Annual Learning for Dynamics and Control Conference}} {\em (\bibinfo{series}{Proceedings of Machine Learning Research})}, Vol.~\bibinfo{volume}{168}. \bibinfo{publisher}{PMLR}, \bibinfo{pages}{263--277}.
\newblock
\showURL{%
\url{https://proceedings.mlr.press/v168/djeumou22a.html}}


\bibitem[\protect\citeauthoryear{Dong, Loy, He, and Tang}{Dong et~al\mbox{.}}{2015}]%
        {dong2015image}
\bibfield{author}{\bibinfo{person}{Chao Dong}, \bibinfo{person}{Chen~Change Loy}, \bibinfo{person}{Kaiming He}, {and} \bibinfo{person}{Xiaoou Tang}.} \bibinfo{year}{2015}\natexlab{}.
\newblock \showarticletitle{Image super-resolution using deep convolutional networks}.
\newblock \bibinfo{journal}{{\em IEEE transactions on pattern analysis and machine intelligence\/}} \bibinfo{volume}{38}, \bibinfo{number}{2} (\bibinfo{year}{2015}), \bibinfo{pages}{295--307}.
\newblock


\bibitem[\protect\citeauthoryear{Draper-Gil, Lashkari, Mamun, and Ghorbani}{Draper-Gil et~al\mbox{.}}{2016}]%
        {draper2016characterization}
\bibfield{author}{\bibinfo{person}{Gerard Draper-Gil}, \bibinfo{person}{Arash~Habibi Lashkari}, \bibinfo{person}{Mohammad Saiful~Islam Mamun}, {and} \bibinfo{person}{Ali~A Ghorbani}.} \bibinfo{year}{2016}\natexlab{}.
\newblock \showarticletitle{Characterization of encrypted and vpn traffic using time-related}. In \bibinfo{booktitle}{{\em Proceedings of the 2nd international conference on information systems security and privacy (ICISSP)}}. \bibinfo{pages}{407--414}.
\newblock


\bibitem[\protect\citeauthoryear{Fedor, Schoffstall, Davin, and Case}{Fedor et~al\mbox{.}}{1990}]%
        {snmp1}
\bibfield{author}{\bibinfo{person}{Mark Fedor}, \bibinfo{person}{Martin~Lee Schoffstall}, \bibinfo{person}{James~R. Davin}, {and} \bibinfo{person}{Dr. Jeff~D. Case}.} \bibinfo{year}{1990}\natexlab{}.
\newblock \bibinfo{title}{{Simple Network Management Protocol (SNMP)}}.
\newblock \bibinfo{howpublished}{RFC 1157}.   (\bibinfo{date}{May} \bibinfo{year}{1990}).
\newblock


\bibitem[\protect\citeauthoryear{Geng, Liu, Yin, Naik, Prabhakar, Rosenblum, and Vahdat}{Geng et~al\mbox{.}}{2019}]%
        {geng2019simon}
\bibfield{author}{\bibinfo{person}{Yilong Geng}, \bibinfo{person}{Shiyu Liu}, \bibinfo{person}{Zi Yin}, \bibinfo{person}{Ashish Naik}, \bibinfo{person}{Balaji Prabhakar}, \bibinfo{person}{Mendel Rosenblum}, {and} \bibinfo{person}{Amin Vahdat}.} \bibinfo{year}{2019}\natexlab{}.
\newblock \showarticletitle{{SIMON: A Simple and Scalable Method for Sensing, Inference and Measurement in Data Center Networks}}. In \bibinfo{booktitle}{{\em 16th USENIX Symposium on Networked Systems Design and Implementation (NSDI 19)}}. \bibinfo{pages}{549--564}.
\newblock


\bibitem[\protect\citeauthoryear{Ghabashneh, Zhao, Lumezanu, Spring, Sundaresan, and Rao}{Ghabashneh et~al\mbox{.}}{2022}]%
        {ghab2022millisamp}
\bibfield{author}{\bibinfo{person}{Ehab Ghabashneh}, \bibinfo{person}{Yimeng Zhao}, \bibinfo{person}{Cristian Lumezanu}, \bibinfo{person}{Neil Spring}, \bibinfo{person}{Srikanth Sundaresan}, {and} \bibinfo{person}{Sanjay Rao}.} \bibinfo{year}{2022}\natexlab{}.
\newblock \showarticletitle{A microscopic view of bursts, buffer contention, and loss in data centers}. In \bibinfo{booktitle}{{\em Proceedings of the 22nd ACM Internet Measurement Conference}} {\em (\bibinfo{series}{IMC '22})}. \bibinfo{publisher}{Association for Computing Machinery}, \bibinfo{address}{New York, NY, USA}, \bibinfo{pages}{567–580}.
\newblock
\showISBNx{9781450392594}
\showDOI{%
\url{https://doi.org/10.1145/3517745.3561430}}


\bibitem[\protect\citeauthoryear{Gill, Diot, Ohlsen, Mathis, and Soltesz}{Gill et~al\mbox{.}}{2022}]%
        {phillipa2022mlab}
\bibfield{author}{\bibinfo{person}{Phillipa Gill}, \bibinfo{person}{Christophe Diot}, \bibinfo{person}{Lai~Yi Ohlsen}, \bibinfo{person}{Matt Mathis}, {and} \bibinfo{person}{Stephen Soltesz}.} \bibinfo{year}{2022}\natexlab{}.
\newblock \showarticletitle{M-Lab: User initiated Internet data for the research community}.
\newblock \bibinfo{journal}{{\em ACM SIGCOMM Computer Communication Review\/}} \bibinfo{volume}{52}, \bibinfo{number}{1}, \bibinfo{pages}{34--37}.
\newblock


\bibitem[\protect\citeauthoryear{Houidi, Azorin, Gallo, Finamore, and Rossi}{Houidi et~al\mbox{.}}{2022}]%
        {houidi2022represent}
\bibfield{author}{\bibinfo{person}{Zied~Ben Houidi}, \bibinfo{person}{Raphael Azorin}, \bibinfo{person}{Massimo Gallo}, \bibinfo{person}{Alessandro Finamore}, {and} \bibinfo{person}{Dario Rossi}.} \bibinfo{year}{2022}\natexlab{}.
\newblock \showarticletitle{{Towards a Systematic Multi-Modal Representation Learning for Network Data}}. In \bibinfo{booktitle}{{\em Proceedings of the 21st ACM Workshop on Hot Topics in Networks}}. \bibinfo{pages}{181–187}.
\newblock


\bibitem[\protect\citeauthoryear{Jiang, Liu, Gember-Jacobson, Schmitt, Bronzino, and Feamster}{Jiang et~al\mbox{.}}{2023}]%
        {jiang2023diffusion}
\bibfield{author}{\bibinfo{person}{Xi Jiang}, \bibinfo{person}{Shinan Liu}, \bibinfo{person}{Aaron Gember-Jacobson}, \bibinfo{person}{Paul Schmitt}, \bibinfo{person}{Francesco Bronzino}, {and} \bibinfo{person}{Nick Feamster}.} \bibinfo{year}{2023}\natexlab{}.
\newblock \showarticletitle{Generative, High-Fidelity Network Traces}. In \bibinfo{booktitle}{{\em Proceedings of the 22nd ACM Workshop on Hot Topics in Networks}} {\em (\bibinfo{series}{HotNets '23})}. \bibinfo{publisher}{Association for Computing Machinery}, \bibinfo{address}{New York, NY, USA}, \bibinfo{pages}{131–138}.
\newblock
\showISBNx{9798400704154}
\showDOI{%
\url{https://doi.org/10.1145/3626111.3628196}}


\bibitem[\protect\citeauthoryear{Karniadakis, Kevrekidis, Lu, Perdikaris, Wang, and Yang}{Karniadakis et~al\mbox{.}}{2021}]%
        {pinn}
\bibfield{author}{\bibinfo{person}{George~Em Karniadakis}, \bibinfo{person}{Ioannis~G Kevrekidis}, \bibinfo{person}{Lu Lu}, \bibinfo{person}{Paris Perdikaris}, \bibinfo{person}{Sifan Wang}, {and} \bibinfo{person}{Liu Yang}.} \bibinfo{year}{2021}\natexlab{}.
\newblock \showarticletitle{Physics-informed machine learning}.
\newblock \bibinfo{journal}{{\em Nature Reviews Physics\/}} \bibinfo{volume}{3}, \bibinfo{number}{6} (\bibinfo{year}{2021}), \bibinfo{pages}{422--440}.
\newblock


\bibitem[\protect\citeauthoryear{Kuzmanovic and Knightly}{Kuzmanovic and Knightly}{2003}]%
        {shrew}
\bibfield{author}{\bibinfo{person}{Aleksandar Kuzmanovic} {and} \bibinfo{person}{Edward~W Knightly}.} \bibinfo{year}{2003}\natexlab{}.
\newblock \showarticletitle{Low-rate TCP-targeted denial of service attacks: the shrew vs. the mice and elephants}. In \bibinfo{booktitle}{{\em Proceedings of the ACM SIGCOMM 2003 conference}}. \bibinfo{pages}{75--86}.
\newblock


\bibitem[\protect\citeauthoryear{Langlet, Ben~Basat, Oliaro, Mitzenmacher, Yu, and Antichi}{Langlet et~al\mbox{.}}{2023}]%
        {langlet2023dta}
\bibfield{author}{\bibinfo{person}{Jonatan Langlet}, \bibinfo{person}{Ran Ben~Basat}, \bibinfo{person}{Gabriele Oliaro}, \bibinfo{person}{Michael Mitzenmacher}, \bibinfo{person}{Minlan Yu}, {and} \bibinfo{person}{Gianni Antichi}.} \bibinfo{year}{2023}\natexlab{}.
\newblock \showarticletitle{Direct Telemetry Access}. In \bibinfo{booktitle}{{\em Proceedings of the ACM SIGCOMM 2023 Conference}} {\em (\bibinfo{series}{ACM SIGCOMM '23})}. \bibinfo{publisher}{Association for Computing Machinery}, \bibinfo{address}{New York, NY, USA}.
\newblock
\showISBNx{9798400702365}
\showDOI{%
\url{https://doi.org/10.1145/3603269.3604827}}


\bibitem[\protect\citeauthoryear{Le, Srivatsa, Ganti, and Sekar}{Le et~al\mbox{.}}{2022}]%
        {le2022foundation}
\bibfield{author}{\bibinfo{person}{Franck Le}, \bibinfo{person}{Mudhakar Srivatsa}, \bibinfo{person}{Raghu Ganti}, {and} \bibinfo{person}{Vyas Sekar}.} \bibinfo{year}{2022}\natexlab{}.
\newblock \showarticletitle{{Rethinking Data-Driven Networking with Foundation Models: Challenges and Opportunities}}. In \bibinfo{booktitle}{{\em Proceedings of the 21st ACM Workshop on Hot Topics in Networks}}. \bibinfo{pages}{188–197}.
\newblock


\bibitem[\protect\citeauthoryear{Ledig, Theis, Huszar, Caballero, Cunningham, Acosta, Aitken, Tejani, Totz, Wang, and Shi}{Ledig et~al\mbox{.}}{2017}]%
        {ledig2017photo}
\bibfield{author}{\bibinfo{person}{Christian Ledig}, \bibinfo{person}{Lucas Theis}, \bibinfo{person}{Ferenc Huszar}, \bibinfo{person}{Jose Caballero}, \bibinfo{person}{Andrew Cunningham}, \bibinfo{person}{Alejandro Acosta}, \bibinfo{person}{Andrew Aitken}, \bibinfo{person}{Alykhan Tejani}, \bibinfo{person}{Johannes Totz}, \bibinfo{person}{Zehan Wang}, {and} \bibinfo{person}{Wenzhe Shi}.} \bibinfo{year}{2017}\natexlab{}.
\newblock \showarticletitle{{Photo-Realistic Single Image Super-Resolution Using a Generative Adversarial Network}}. In \bibinfo{booktitle}{{\em Proceedings of the IEEE Conference on Computer Vision and Pattern Recognition (CVPR)}}.
\newblock


\bibitem[\protect\citeauthoryear{Lei, Yu, Liu, and Xu}{Lei et~al\mbox{.}}{2022}]%
        {lei2022printq}
\bibfield{author}{\bibinfo{person}{Yiran Lei}, \bibinfo{person}{Liangcheng Yu}, \bibinfo{person}{Vincent Liu}, {and} \bibinfo{person}{Mingwei Xu}.} \bibinfo{year}{2022}\natexlab{}.
\newblock \showarticletitle{PrintQueue: performance diagnosis via queue measurement in the data plane}. In \bibinfo{booktitle}{{\em Proceedings of the ACM SIGCOMM 2022 Conference}} {\em (\bibinfo{series}{SIGCOMM '22})}. \bibinfo{publisher}{Association for Computing Machinery}, \bibinfo{address}{New York, NY, USA}, \bibinfo{pages}{516–529}.
\newblock
\showISBNx{9781450394208}
\showDOI{%
\url{https://doi.org/10.1145/3544216.3544257}}


\bibitem[\protect\citeauthoryear{Li, Miao, Kim, and Yu}{Li et~al\mbox{.}}{2016}]%
        {li2016flowradar}
\bibfield{author}{\bibinfo{person}{Yuliang Li}, \bibinfo{person}{Rui Miao}, \bibinfo{person}{Changhoon Kim}, {and} \bibinfo{person}{Minlan Yu}.} \bibinfo{year}{2016}\natexlab{}.
\newblock \showarticletitle{$\{$FlowRadar$\}$: A Better $\{$NetFlow$\}$ for Data Centers}. In \bibinfo{booktitle}{{\em 13th USENIX symposium on networked systems design and implementation (NSDI 16)}}. \bibinfo{pages}{311--324}.
\newblock


\bibitem[\protect\citeauthoryear{Lin, Jain, Wang, Fanti, and Sekar}{Lin et~al\mbox{.}}{2020}]%
        {lin2020dp}
\bibfield{author}{\bibinfo{person}{Zinan Lin}, \bibinfo{person}{Alankar Jain}, \bibinfo{person}{Chen Wang}, \bibinfo{person}{Giulia Fanti}, {and} \bibinfo{person}{Vyas Sekar}.} \bibinfo{year}{2020}\natexlab{}.
\newblock \showarticletitle{Using gans for sharing networked time series data: Challenges, initial promise, and open questions}. In \bibinfo{booktitle}{{\em Proceedings of the ACM Internet Measurement Conference}}. \bibinfo{pages}{464--483}.
\newblock


\bibitem[\protect\citeauthoryear{NS3}{NS3}{2023}]%
        {ns3}
\bibfield{author}{\bibinfo{person}{NS3}.} \bibinfo{year}{2023}\natexlab{}.
\newblock \showarticletitle{{NS3 Network Simulator}}. \bibinfo{howpublished}{\url{https://www.nsnam.org/}}.
\newblock


\bibitem[\protect\citeauthoryear{Pedregosa, Varoquaux, Gramfort, Michel, Grisel, Blondel, Prettenhofer, Dubourg, Vanderplas, Passos, Cournapeau, Brucher, Perrot, and Duchesnay}{Pedregosa et~al\mbox{.}}{2011}]%
        {scikit-learn}
\bibfield{author}{\bibinfo{person}{F. Pedregosa}, \bibinfo{person}{G. Varoquaux}, \bibinfo{person}{A. Gramfort}, \bibinfo{person}{B. Michel, V.and~Thirion}, \bibinfo{person}{O. Grisel}, \bibinfo{person}{M. Blondel}, \bibinfo{person}{R. Prettenhofer, P.and~Weiss}, \bibinfo{person}{V. Dubourg}, \bibinfo{person}{J. Vanderplas}, \bibinfo{person}{A. Passos}, \bibinfo{person}{D. Cournapeau}, \bibinfo{person}{M. Brucher}, \bibinfo{person}{M. Perrot}, {and} \bibinfo{person}{E. Duchesnay}.} \bibinfo{year}{2011}\natexlab{}.
\newblock \showarticletitle{Scikit-learn: Machine Learning in {P}ython}.
\newblock \bibinfo{journal}{{\em Journal of Machine Learning Research\/}}  \bibinfo{volume}{12} (\bibinfo{year}{2011}), \bibinfo{pages}{2825--2830}.
\newblock


\bibitem[\protect\citeauthoryear{Qian, Cui, Tso, Deng, and Jia}{Qian et~al\mbox{.}}{2023}]%
        {qian2023offset}
\bibfield{author}{\bibinfo{person}{Mimi Qian}, \bibinfo{person}{Lin Cui}, \bibinfo{person}{Fung~Po Tso}, \bibinfo{person}{Yuhui Deng}, {and} \bibinfo{person}{Weijia Jia}.} \bibinfo{year}{2023}\natexlab{}.
\newblock \showarticletitle{OffsetINT: Achieving High Accuracy and Low Bandwidth for In-Band Network Telemetry}.
\newblock \bibinfo{journal}{{\em IEEE Transactions on Services Computing\/}} (\bibinfo{year}{2023}), \bibinfo{pages}{1--12}.
\newblock
\showDOI{%
\url{https://doi.org/10.1109/TSC.2023.3323697}}


\bibitem[\protect\citeauthoryear{Rasley, Stephens, Dixon, Rozner, Felter, Agarwal, Carter, and Fonseca}{Rasley et~al\mbox{.}}{2014}]%
        {rasley2014planck}
\bibfield{author}{\bibinfo{person}{Jeff Rasley}, \bibinfo{person}{Brent Stephens}, \bibinfo{person}{Colin Dixon}, \bibinfo{person}{Eric Rozner}, \bibinfo{person}{Wes Felter}, \bibinfo{person}{Kanak Agarwal}, \bibinfo{person}{John Carter}, {and} \bibinfo{person}{Rodrigo Fonseca}.} \bibinfo{year}{2014}\natexlab{}.
\newblock \showarticletitle{Planck: Millisecond-scale monitoring and control for commodity networks}.
\newblock \bibinfo{journal}{{\em ACM SIGCOMM Computer Communication Review\/}} \bibinfo{volume}{44}, \bibinfo{number}{4} (\bibinfo{year}{2014}), \bibinfo{pages}{407--418}.
\newblock


\bibitem[\protect\citeauthoryear{Sengupta, Kim, and Rexford}{Sengupta et~al\mbox{.}}{2022}]%
        {sengupta2022continuous}
\bibfield{author}{\bibinfo{person}{Satadal Sengupta}, \bibinfo{person}{Hyojoon Kim}, {and} \bibinfo{person}{Jennifer Rexford}.} \bibinfo{year}{2022}\natexlab{}.
\newblock \showarticletitle{Continuous In-Network Round-Trip Time Monitoring}. In \bibinfo{booktitle}{{\em Proceedings of the ACM SIGCOMM 2022 Conference}}. \bibinfo{address}{New York, NY, USA}, \bibinfo{pages}{473–485}.
\newblock
\showDOI{%
\url{https://doi.org/10.1145/3544216.3544222}}


\bibitem[\protect\citeauthoryear{Teixeira, Shaikh, Griffin, and Rexford}{Teixeira et~al\mbox{.}}{2008}]%
        {teixeira2008impact}
\bibfield{author}{\bibinfo{person}{Renata Teixeira}, \bibinfo{person}{Aman Shaikh}, \bibinfo{person}{Timothy~G Griffin}, {and} \bibinfo{person}{Jennifer Rexford}.} \bibinfo{year}{2008}\natexlab{}.
\newblock \showarticletitle{Impact of hot-potato routing changes in IP networks}.
\newblock \bibinfo{journal}{{\em IEEE/ACM Transactions On Networking\/}} \bibinfo{volume}{16}, \bibinfo{number}{6} (\bibinfo{year}{2008}), \bibinfo{pages}{1295--1307}.
\newblock


\bibitem[\protect\citeauthoryear{Tilmans, B{\"u}hler, Poese, Vissicchio, and Vanbever}{Tilmans et~al\mbox{.}}{2018}]%
        {tilmans2018stroboscope}
\bibfield{author}{\bibinfo{person}{Olivier Tilmans}, \bibinfo{person}{Tobias B{\"u}hler}, \bibinfo{person}{Ingmar Poese}, \bibinfo{person}{Stefano Vissicchio}, {and} \bibinfo{person}{Laurent Vanbever}.} \bibinfo{year}{2018}\natexlab{}.
\newblock \showarticletitle{Stroboscope: Declarative network monitoring on a budget}. In \bibinfo{booktitle}{{\em 15th USENIX Symposium on Networked Systems Design and Implementation (NSDI 18)}}. \bibinfo{pages}{467--482}.
\newblock


\bibitem[\protect\citeauthoryear{Vaswani, Shazeer, Parmar, Uszkoreit, Jones, Gomez, Kaiser, and Polosukhin}{Vaswani et~al\mbox{.}}{2017}]%
        {vaswani2017attention}
\bibfield{author}{\bibinfo{person}{Ashish Vaswani}, \bibinfo{person}{Noam Shazeer}, \bibinfo{person}{Niki Parmar}, \bibinfo{person}{Jakob Uszkoreit}, \bibinfo{person}{Llion Jones}, \bibinfo{person}{Aidan~N Gomez}, \bibinfo{person}{\L~ukasz Kaiser}, {and} \bibinfo{person}{Illia Polosukhin}.} \bibinfo{year}{2017}\natexlab{}.
\newblock \showarticletitle{{Attention is All you Need}}. In \bibinfo{booktitle}{{\em Advances in Neural Information Processing Systems}}, \bibfield{editor}{\bibinfo{person}{I.~Guyon}, \bibinfo{person}{U.~Von Luxburg}, \bibinfo{person}{S.~Bengio}, \bibinfo{person}{H.~Wallach}, \bibinfo{person}{R.~Fergus}, \bibinfo{person}{S.~Vishwanathan}, {and} \bibinfo{person}{R.~Garnett}} (Eds.), Vol.~\bibinfo{volume}{30}. \bibinfo{publisher}{Curran Associates, Inc.}
\newblock
\showURL{%
\url{https://proceedings.neurips.cc/paper_files/paper/2017/file/3f5ee243547dee91fbd053c1c4a845aa-Paper.pdf}}


\bibitem[\protect\citeauthoryear{Wang, Xie, Dong, and Shan}{Wang et~al\mbox{.}}{2021}]%
        {wang2021real}
\bibfield{author}{\bibinfo{person}{Xintao Wang}, \bibinfo{person}{Liangbin Xie}, \bibinfo{person}{Chao Dong}, {and} \bibinfo{person}{Ying Shan}.} \bibinfo{year}{2021}\natexlab{}.
\newblock \showarticletitle{{Real-esrgan: Training real-world blind super-resolution with pure synthetic data}}. In \bibinfo{booktitle}{{\em Proceedings of the IEEE/CVF International Conference on Computer Vision}}. \bibinfo{pages}{1905--1914}.
\newblock


\bibitem[\protect\citeauthoryear{Wang, Yu, Wu, Gu, Liu, Dong, Qiao, and Change~Loy}{Wang et~al\mbox{.}}{2018}]%
        {wang2018esrgan}
\bibfield{author}{\bibinfo{person}{Xintao Wang}, \bibinfo{person}{Ke Yu}, \bibinfo{person}{Shixiang Wu}, \bibinfo{person}{Jinjin Gu}, \bibinfo{person}{Yihao Liu}, \bibinfo{person}{Chao Dong}, \bibinfo{person}{Yu Qiao}, {and} \bibinfo{person}{Chen Change~Loy}.} \bibinfo{year}{2018}\natexlab{}.
\newblock \showarticletitle{{Esrgan: Enhanced super-resolution generative adversarial networks}}. In \bibinfo{booktitle}{{\em Proceedings of the European conference on computer vision (ECCV) workshops}}. \bibinfo{pages}{0--0}.
\newblock


\bibitem[\protect\citeauthoryear{Wu, Li, Yang, Zhang, Sheng, and Hou}{Wu et~al\mbox{.}}{2023}]%
        {wu2023learning}
\bibfield{author}{\bibinfo{person}{Ming Wu}, \bibinfo{person}{Qianmu Li}, \bibinfo{person}{Fei Yang}, \bibinfo{person}{Jing Zhang}, \bibinfo{person}{Victor~S Sheng}, {and} \bibinfo{person}{Jun Hou}.} \bibinfo{year}{2023}\natexlab{}.
\newblock \showarticletitle{Learning from biased crowdsourced labeling with deep clustering}.
\newblock \bibinfo{journal}{{\em Expert Systems with Applications\/}}  \bibinfo{volume}{211} (\bibinfo{year}{2023}), \bibinfo{pages}{118608}.
\newblock


\bibitem[\protect\citeauthoryear{Wu, Li, Zhang, and Hou}{Wu et~al\mbox{.}}{2022}]%
        {wu2022label}
\bibfield{author}{\bibinfo{person}{Ming Wu}, \bibinfo{person}{Qianmu Li}, \bibinfo{person}{Jing Zhang}, {and} \bibinfo{person}{Jun Hou}.} \bibinfo{year}{2022}\natexlab{}.
\newblock \showarticletitle{Label Aggregation with Clustering for Biased Crowdsourced Labeling}. In \bibinfo{booktitle}{{\em Proceedings of the 2022 14th International Conference on Machine Learning and Computing}} {\em (\bibinfo{series}{ICMLC '22})}. \bibinfo{publisher}{Association for Computing Machinery}, \bibinfo{address}{New York, NY, USA}, \bibinfo{pages}{165–169}.
\newblock
\showISBNx{9781450395700}
\showDOI{%
\url{https://doi.org/10.1145/3529836.3529861}}


\bibitem[\protect\citeauthoryear{Yin, Lin, Jin, Fanti, and Sekar}{Yin et~al\mbox{.}}{2022}]%
        {yin2022netshare}
\bibfield{author}{\bibinfo{person}{Yucheng Yin}, \bibinfo{person}{Zinan Lin}, \bibinfo{person}{Minhao Jin}, \bibinfo{person}{Giulia Fanti}, {and} \bibinfo{person}{Vyas Sekar}.} \bibinfo{year}{2022}\natexlab{}.
\newblock \showarticletitle{Practical GAN-based synthetic IP header trace generation using NetShare}. In \bibinfo{booktitle}{{\em Proceedings of the ACM SIGCOMM 2022 Conference}} {\em (\bibinfo{series}{SIGCOMM '22})}. \bibinfo{publisher}{Association for Computing Machinery}, \bibinfo{address}{New York, NY, USA}.
\newblock
\showISBNx{9781450394208}
\showDOI{%
\url{https://doi.org/10.1145/3544216.3544251}}


\bibitem[\protect\citeauthoryear{Zhang, Gool, and Timofte}{Zhang et~al\mbox{.}}{2020}]%
        {zhang2020deep}
\bibfield{author}{\bibinfo{person}{Kai Zhang}, \bibinfo{person}{Luc~Van Gool}, {and} \bibinfo{person}{Radu Timofte}.} \bibinfo{year}{2020}\natexlab{}.
\newblock \showarticletitle{Deep unfolding network for image super-resolution}. In \bibinfo{booktitle}{{\em Proceedings of the IEEE/CVF conference on computer vision and pattern recognition}}. \bibinfo{pages}{3217--3226}.
\newblock


\bibitem[\protect\citeauthoryear{Zhang, Tian, Kong, Zhong, and Fu}{Zhang et~al\mbox{.}}{2018}]%
        {zhang2018residual}
\bibfield{author}{\bibinfo{person}{Yulun Zhang}, \bibinfo{person}{Yapeng Tian}, \bibinfo{person}{Yu Kong}, \bibinfo{person}{Bineng Zhong}, {and} \bibinfo{person}{Yun Fu}.} \bibinfo{year}{2018}\natexlab{}.
\newblock \showarticletitle{Residual dense network for image super-resolution}. In \bibinfo{booktitle}{{\em Proceedings of the IEEE conference on computer vision and pattern recognition}}. \bibinfo{pages}{2472--2481}.
\newblock


\end{thebibliography}
